\documentstyle[11pt,aaspp]{article}

\makeatletter
\let\@internalcite\cite
\def\cite{\def\citename##1{##1}\@internalcite}
\def\shortcite{\def\citename##1{}\@internalcite}
\makeatother

\newcommand{\constant}{\mbox{constant}}
\newcommand{\einternal}{e_{\it int\/}}
\newcommand{\etotal}{e_{\it tot\/}}

\newcommand{\Lbar}{\overline{L}}

\newcommand{\fone}{f^{(1)}}
\newcommand{\ftwo}{f^{(2)}}
\newcommand{\gone}{g^{(1)}}
\newcommand{\gtwo}{g^{(2)}}

\newcommand{\setofv}{\{v_j\}}
\newcommand{\setofvt}{\{\tilde{v_j}\}}

\newcommand{\pderiv}[2]{\frac{\partial #1}{\partial #2}}

\newcounter{figurenumber}
\newcounter{tempcounter}
\newcommand{\newfigure}{\stepcounter{figurenumber}\mbox{\thefigurenumber}}
\newcommand{\samefigure}{\mbox{\thefigurenumber}}
\newcommand{\prevfigure}{\setcounter{tempcounter}{\value{figurenumber}}%
\addtocounter{tempcounter}{-1}\thetempcounter}
\newcommand{\backfigure}[1]{\setcounter{tempcounter}{\value{figurenumber}}%
\addtocounter{tempcounter}{-#1}\thetempcounter}
\newcommand{\twobackfigure}{\setcounter{tempcounter}{\value{figurenumber}}%
\addtocounter{tempcounter}{-2}\thetempcounter}

\newcounter{sect}
\newcommand{\nextsect}{\stepcounter{sect}\mbox{\thesect}}

\newenvironment{captions}
  {
   \begin{list}{??}{
                    \setlength{\leftmargin}{\captlabelwidth}
                    \setlength{\rightmargin}{0em}
                   }
  }{\end{list}}
\newlength{\captlabelwidth}
\begin{document}

\title{Convergence~Properties of Finite-Difference~Hydrodynamics Schemes
        in~the~Presence of~Shocks}
\author{Paul A. Kimoto}
\affil{Department of Physics, Cornell University, Ithaca, NY 14853}

\and 

\author{David F. Chernoff}
\affil{Department of Astronomy, Cornell University, Ithaca, NY 14853}



\begin{abstract}

\noindent
We investigate asymptotic convergence in the~$\Delta x \!\rightarrow\! 0$ limit
as a tool for determining whether numerical computations involving shocks are
accurate.  We use one-dimensional operator-split finite-difference schemes for
hydrodynamics with a von Neumann artificial viscosity.  An internal-energy
scheme converges to demonstrably wrong solutions.  We associate this failure
with the presence of discontinuities in the limiting solution.  Our extension
of the Lax-Wendroff theorem guarantees that certain conservative,
operator-split schemes converge to the correct continuum solution.  For such a
total-energy scheme applied to the formation of a single shock, convergence of
a Cauchy error approaches the expected rate slowly.  We relate this slowness to
the effect of varying diffusion, due to varying linear artificial-viscous
length, on small-amplitude waves.  In an appendix we discuss the scaling of
shock-transition regions with viscous lengths, and exhibit several difficulties
for attempts to make extrapolations.

\end{abstract}



\section{Introduction}

Finite-difference hydrodynamics schemes find widespread use in astrophysical
applications, where they are often parts of computations that involve other
significant physical processes, such as external heating and radiative cooling.
In this paper we consider numerical methods for establishing the accuracy of a
calculation in the presence of shocks.  Our primary tool is testing for
convergence.  We require that calculations not merely converge, but converge at
or near the expected asymptotic rate.  This accuracy criterion should readily
lend itself to automated refinement methods in which grids are made denser only
when and where required for accuracy.

For our investigations we use the one-dimensional Euler equations modified by
the inclusion of an explicit artificial viscosity.  The numerical schemes are
operator-split and explicit.  We describe two schemes: a {\em
non-conservative\/} form that uses the internal-energy density as a primary
variable, as is common in astrophysical calculations (e.g., \cite{nw},
\cite{stone}, and \cite{hsw}), and a similar {\em conservative\/} form that
instead uses the total-energy density as a primary variable.

\setcounter{sect}{\value{section}}

In this paper we investigate the performance of these schemes in simple test
problems with shocks.  After outlining the numerical methods in
section~\nextsect , we show in section~\nextsect\ that the non-conservative
internal-energy scheme converges to incorrect solutions when shocks develop.
This failure occurs when artificial-viscous lengths and gridsize are decreased
proportionally, as is the usual practice.  When viscous lengths are held fixed
as the gridsize tends to zero, however, the scheme converges to a correct
solution (albeit to a different physical problem).  Section~\nextsect\
discusses the performance of both schemes in a shock-tube problem.  We show
explicitly that the internal-energy scheme (for a particular choice of viscous
lengths) converges toward a solution with shock speed too low by
approximately~0.2\%.  The other, conservative difference scheme converges
toward the correct solution even in the presence of the shock.  (We show in
Appendix~A that its correct behavior is a consequence of the Lax-Wendroff
theorem~[\cite{lw}], which we extend to operator-split schemes.)

Then in section~\nextsect\ we examine rates of convergence for the conservative
scheme.  We show that the presence of linear artificial viscosity (proportional
to~$\partial v/\partial x$) introduces a substantial amount of diffusion.  In
sequences in which the artificial viscosity vanishes with the gridsize, the
variation in diffusion has an observable effect on convergence rates for
small-amplitude waves.  Finally, we consider a wall-shock problem, which also
contains such small-amplitude waves.  We consider two kinds of sequences:
(1)~sequences in which the artificial viscosity enters as a constant term, and
(2)~sequences in which it varies with the gridsize.  Holding the
artificial-viscous lengths fixed leads quite readily to small changes and the
expected convergence rate.  In contrast, in the sequences where the viscosity
vanishes with the gridsize, we observe that the convergence rate tends only
slowly toward the rate expected.  The effect of linear artificial viscosity on
post-shock sound waves contributes substantially to this slowness.

Attempts to extrapolate solutions with different viscous lengths must cope with
the variation in the widths of transitions that represent shocks.  These widths
closely scale with the artificial-viscous lengths, but small deviations from
exact scaling behavior may pose considerable barriers to extrapolation
algorithms.  These deviations from scaling are the main subject of Appendix~C.


\section{Numerical Methods}

We refer to our two finite-difference schemes as ``internal-energy'' and
``total-energy,'' after the underlying differential equations that they
represent.  We consider the one-dimensional Euler equations, written in the
form
\begin{eqnarray}
  \label{euler_density}
    \frac{\partial \rho}{\partial t} & = & -\frac{\partial m}{\partial x} \\
  \label{euler_momentum}
    \frac{\partial m}{\partial t} &
          = & -\frac{\partial}{\partial x} \left[\frac{m^2}{\rho} + P\right] \\
  \label{euler_internal_energy}
    \frac{\partial \einternal}{\partial t} &
      = & -\left(\frac{\partial}{\partial x}
                          \left[\frac{\einternal \,m}{\rho}\right]
            + P \frac{\partial}{\partial x} \frac{m}{\rho}\right) \\
  \label{euler_total_energy}
    \frac{\partial \etotal}{\partial t} &
               = & -\frac{\partial}{\partial x}
                               \left[(\etotal + P)\frac{m}{\rho}\right],
\end{eqnarray}
where $m = \rho v\/$ is the momentum density (here per length); $\einternal$,
the {\em internal\/} energy density; $\etotal$, the {\em total\/} energy
density, and $P\/$, the pressure.  The equation of
state is given by the ideal gas law; generally we take~$\gamma = 5/3$.
Equations (\ref{euler_internal_energy}) and (\ref{euler_total_energy}) are
alternative expressions of the conservation of energy.  In the internal-energy
scheme, we apply equation~(\ref{euler_internal_energy}); in the
total-energy scheme, equation~(\ref{euler_total_energy}).

Since we want to capture shocks automatically and so avoid the
computational difficulties of shock tracking, we use an explicit artificial
viscosity to generate the entropy necessary for shocks.  Following Norman and
Winkler~(\shortcite{nw}), to the physical pressure~$P$ we add the term
\begin{equation}             \label{av}
   Q = \rho \frac{\partial v}{\partial x}
            \left[-L_1 c
                 + L_2^2 \min\left(\frac{\partial v}{\partial x},
                                                        0\right)\right],
\end{equation}
where $c\/$ is the adiabatic sound speed.  The constants~$L_1$ and~$L_2$ are
viscous lengths for linear and quadratic artificial-viscous terms, and
parametrize their strengths.  Typically the artificial viscosity smears shock
transitions over a region whose size is comparable to the viscous lengths.
Conventionally the viscous lengths~$L_1$ and~$L_2$ are chosen to be a small
multiple of the gridsize~$\Delta x$; we adopt this approach for our initial
investigations.

For a finite-difference representation of the equations of motion, we employ an
operator-split scheme similar to the internal-energy scheme developed by Norman
and Winkler~(\cite{nw}, \cite{stone}).  They use a staggered spatial grid,
defining scalar and vector quantities at different gridpoints; whereas our grid
is unstaggered.

The operator splitting proceeds as follows: for the parts of the
equations~(\ref{euler_density}--\ref{euler_total_energy}) that represent
advection,
\begin{equation}         \label{advection}
  \frac{\partial \psi}{\partial t} = -\frac{\partial [v \psi]}{\partial x},
\end{equation}
we use the monotonic advection scheme of van Leer~(\shortcite{vanleer}).  The
rest of the terms are
treated in three further substeps: pressure acceleration, artificial viscosity,
and compressional heating.  In each full timestep, we account first for each of
these three terms (in the order specified), and then for the advection terms.
Each substep generates new values of the physical variables, which then are
used in the update for the next substep.

Explicitly, then, the finite-difference equations take this form: first we
represent the pressure-acceleration term by
\begin{equation}      \label{pressure_acceleration}
  m_j \mapsto 
    m_j - \frac{\Delta t}{2\Delta x} \left( P_{j+1} - P_{j-1} \right) .
\end{equation}
To incorporate the artificial-viscosity terms we next use
\begin{eqnarray}
  m_j & \mapsto &              \label{art-visc-momentum}
     m_j - \frac{\Delta t}{2\Delta x}\left(Q_{j+1} - Q_{j-1} \right) \\
                               \label{art-visc-internal}
  (\einternal)_j & \mapsto &
   (\einternal)_j - \frac{\Delta t}{2\Delta x}
                  Q_j \left(v_{j+1} - v_{j-1} \right)
              \label{av-difference} \\
  (\etotal)_j & \mapsto &
   (\etotal)_j - \frac{\Delta t}{2\Delta x} (Q_{j+1}v_{j+1} - Q_{j-1}v_{j-1}) ,
\end{eqnarray}
where values of the momentum density---updated by pressure
acceleration---enter into the right-hand sides through~$v_j = m_j/\rho_j$
and~$Q_j = \rho_j (\partial v/\partial x)_j
[ -L_1 c_j + L_2^2 \min\left( 0, (\partial v/\partial x)_j \right) ]$.
For the gradient term, we take~$(\partial v/\partial x)_j = (v_{j+1} -
v_{j-1})/2\Delta x$.
Then, in the compressional-heating substep, the energy density is changed by
\begin{eqnarray}        \label{comp-heat-difference}
  (\einternal)_j  & \mapsto &
     (\einternal)_j - \frac{\Delta t}{2\Delta x} (\einternal)_j
                            (v_{j+1} - v_{j-1}) \\
  (\etotal)_j & \mapsto &
     (\etotal)_j
      - \frac{\Delta t}{2\Delta x}
          \left[ (\etotal)_{j+1}v_{j+1} - (\etotal)_{j-1}v_{j-1} \right],
\end{eqnarray}
where once again the latest values of~$(\einternal)_j$,~$(\etotal)_j$,
and~$m_j$ enter the right-hand sides.

The resulting values of the physical quantities enter the advection substep.
The van Leer procedure has the flux-conserving form
\begin{equation}
  \psi_j \mapsto
    \psi_j - \frac{\Delta t}{\Delta x} ({\cal F}_{j+1/2} - {\cal F}_{j-1/2}).
\end{equation}
The fluxes~${\cal F}_{j+1/2}$ are calculated in the upwind fashion
\begin{equation}
  {\cal F}_{j+1/2} =
    \left\{
      \begin{array}{ll}
         v_{j+1/2}\left[\psi_j
             - \frac{1}{2}\left(v_{j+1/2}\Delta t/\Delta x - 1\right)
                                     \nabla\psi_{j+1}\right], &
               \mbox{if $ v_{j+1/2} \geq 0 $,} \\
         v_{j+1/2}\left[\psi_{j+1}
          - \frac{1}{2}\left(v_{j+1/2}\Delta t/\Delta x + 1\right)
                                                 \nabla\psi_{j+1}\right], &
               \mbox{if \(v_{j+1/2} < 0\),}
      \end{array}
   \right.
\end{equation}
where $v_{j+1/2} = (v_{j+1} + v_j)/2$, and (as always) $ v_j = m_j/\rho_j$.
This advection scheme's monotonicity arises from van Leer's
definition of the gradient-like term
\begin{equation}
  \nabla \psi_j  =  \left\{
                        \begin{array}{ll}
                          {2\delta\psi_{j+1/2} \delta\psi_{j-1/2} /
                                              (\psi_{j+1} - \psi_{j-1})} &
                           \mbox{if
                             $\delta\psi_{j+1/2} \delta\psi_{j-1/2} > 0$,}\\
                          0 &\mbox{otherwise,}
                        \end{array}
                      \right.
\end{equation}
where \( \delta \psi_{j+1/2} = \psi_{j+1} - \psi_j \).  The spatial accuracy of
the van Leer procedure is second-order, except at extrema, where the accuracy
must be reduced to first order to guarantee monotonicity.  After the
advection substep, the update through the timestep~$\Delta t$ is complete.

In regions where the continuum solution is smooth, such difference schemes are
consistent with the differential equations~(\cite{sod85}).  When
discontinuities are present, however, it is not immediately obvious that
truncation errors should vanish in the~$\Delta x, \Delta t \!\rightarrow\! 0$
limit.  If we regard truncation errors as products of these small
quantities and derivatives of the solution, then clearly it is possible that the
truncation errors may remain finite even in the small-gridsize, small-timestep
limit.

Lax and Wendroff~(\shortcite{lw}) proved that single-step {\em conservative\/}
difference schemes, if stable and convergent, converge to weak solutions of the
corresponding systems of conservation laws.  This theorem can be directly
extended to include operator-split, multi-step conservative schemes as well
(see Appendix~A).  We know of no such theorem or extension for non-conservative
schemes, however.  Below we present results that show that some
non-conservative schemes (namely, our internal-energy type) do {\em not\/}
converge to weak solutions of conservation laws.

The operator-splitting method limits the overall time accuracy to first order,
and the spatial accuracy of the substeps is (almost everywhere) second order.
For smooth flows, then, asymptotically the error per step
is~\( O( (\Delta t)^2, \Delta t(\Delta x)^2) \).

In general, the need for numerical stability requires constraints on the
timestep~$\Delta t$~(\cite{richtmyer}).  We impose two constraints.  First, the
timestep must everywhere satisfy the Courant-Friedrichs-Lewy
condition,
\begin{equation}
   \Delta t \le \frac{\Delta x}{\left|v \right| + c}.
\end{equation}
Second, since the artificial viscosity takes the form of diffusion, we
enforce a ``diffusion-limiting'' condition
\begin{equation}                \label{diff-limit}
   \Delta t \le {\frac{(\Delta x)^2}{L_1 c
                       - 2 L_2^2 \min{(\partial v/\partial x, 0)}}}.
\end{equation}
To ensure these conditions, before making each update we choose the timestep to
be
\begin{equation}       \label{delta-t}
  \Delta t = \min_{\it grid}
              \left[
               \frac{C \Delta x}{\left| v \right| + c},
               \frac{D(\Delta x)^2}
                     {L_1 c - 2 L_2^2 \min{(\partial v/\partial x, 0)}}
              \right],
\end{equation}
where $C < 1$ and $D < 1$ are safety factors.  All of the calculations shown in
this paper use~$C = D = 0.9$.

Despite these stability constraints, we have found that our total-energy scheme
is {\em unstable\/} in the limit~$\Delta x \rightarrow 0$ when~$L_1/\Delta x =
\constant$.  In our experience, any given unstable calculation may be
stabilized by increasing the linear artificial viscous length~$L_1$.  However,
{\em we discourage the use of the total-energy scheme for convergence studies.}
Imposing a diffusion-limiting condition~(\ref{diff-limit}) means that
increasing the artificial viscosity forces smaller timesteps and greater
computational expense.  We use the scheme in the context of this paper to
emphasize that our conclusions regarding the internal-energy scheme are
based on that scheme's non-conservative nature.

In the calculations that we show in this paper, we observe no indications of
instability.  Further, we have duplicated most of the total-energy
calculations with the (likewise conservative) classic Lax-Wendroff
scheme~(\cite{lw}) in two-step form~(\cite{richtmyer}).  (Some Lax-Wendroff
calculations required reduced timesteps for stability, achieved by using a
smaller safety factor~$D$.)  These verify that our results regarding
convergence in the presence of linear artificial viscosity depend on the
viscosity, not on the underlying numerical scheme.


\section{Nonlinear Sound Wave}

Since the internal-energy equation~(\ref{euler_internal_energy}) is not in the
form of a conservation law, finite-difference schemes based on it do not
conserve total energy exactly.  Hence, one test of its
results is to compare the calculated total energy with the total energy in the
initial conditions.  Under some conditions this scheme produces
incorrect results.

For the first test problem we consider a steepening wave.  The initial
conditions superpose a large-amplitude wave on a flat background:
\pagebreak[1]
\begin{eqnarray}          \label{right-moving-wave}
  v(x)    & = & \delta  (x) \nonumber \\
  P(x)    & = & 3/5 + \delta (x)  \\
  \rho(x) & = & 1 + \delta (x), \nonumber
\end{eqnarray}
where
\begin{equation}            \label{modified-gaussian}
  \delta (x) =
    \left\{
      \begin{array}{ll}
        \delta_0 e^{ -\left[ (x - x_0)/\Delta \right]^2
                          - \left[ (x - x_0)/\Delta\right]^4 }, &
                              \mbox{if $ \left| x - x_0 \right| < 2\Delta $} \\
        0, & \mbox{otherwise.}
      \end{array}
    \right.
\end{equation}
For the amplitude we take~$\delta_0 = 0.2$; and for the width
parameter,~$\Delta = 100/13$.  In the small-amplitude limit of the inviscid
equations of motion, this waveform is a purely right-moving wave.  Because of
non-linear effects, it steepens and forms a shock on its leading edge.  (The
grid is large enough that waves from the perturbation do not reach the edges;
hence no energy flows in or out the computational volume.)  The viscous
parameters are~$L_1/\Delta x = 1/2$ and~$L_2/\Delta x = 1$.  We refine the
gridsize~$\Delta x$ and compare the results.

Figure~\newfigure\ illustrates the behavior of the total-energy
error~$\Delta E_{\it total}
$ at various times as a function of gridsize~$\Delta x$.  At early
times~($t \le 20$) the error decreases proportionally to~$\Delta x$, as
anticipated for this first-order accurate scheme.  At late times~($t \ge 28$),
however, at small values of~$\Delta x$ the energy error tends toward {\em
non-zero\/} values.  For these cases, increasing resolution leads not to
greater accuracy, but rather toward some incorrect solution.  The
internal-energy scheme cannot be performing correctly at late times.

The shock forms at $t = 23 \pm 0.5$, estimated as follows.  We repeat the
calculations beginning with the same initial conditions but without artificial
viscosity.  To forestall the onset of numerical instabilities that develop in
the absence of linear artificial viscosity, the timestep safety factor~$C$ is
reduced to~$0.4$.  We use the characteristics moving to the right from each
gridpoint at the speed~$v + c$ to find the time of first crossing, which
corresponds to shock formation.  Close examination of the~\mbox{$t = 24$} curve
in Figure~\samefigure\ suggests that it may deviate from the well-behaved
curves for~\mbox{$t \le 20$}.  Once the shock forms, the non-vanishing energy
error becomes increasingly apparent.  This experiment strongly suggests that
the internal-energy scheme fails when the flow contains a shock.

Before discussing this failure further, we introduce another measure of errors
in computed solutions.  Define the Cauchy error for a solution computed with
gridsize~$\Delta x$ by
\begin{equation}             \label{cauchy}
   \epsilon_{\Delta x} [\psi] =
         \int\limits_{\it grid} \left|\psi_{\Delta x}(x, T)
                          - \psi_{\Delta x / 2}(x, T) \right| \,dx,
\end{equation}
where $\psi_{\Delta x}(x, T)$ is some physical variable (at position~$x$ and
time~$T$) as computed with gridsize~$\Delta x$.  When the numerical
calculations converge to the solution of the continuum system, this error
measure should behave as~$O((\Delta x)^\alpha)$, where~$\alpha$ gives the
leading-order behavior of the error.  For smooth flows, $\alpha$~is simply the
order of accuracy of the scheme.  When discontinuities are present, however,
$\alpha$~depends also on the changes in the computed transition widths~$\ell$,
which should be on the order of the viscous lengths~$L_1$ and~$L_2$.  (We
discuss this issue further in Appendix~C.)  If the jump across the transition
is~$\Delta \psi$, then (on dimensional grounds alone) the shock should
contribute~$\ell \Delta\psi$ (up to numerical constants) to the Cauchy error.
When the viscous lengths~$L_1$ and~$L_2$ are taken to be proportional to the
gridsize~$\Delta x$, the contribution to the Cauchy error should also behave
as~$O(\Delta x)$.  (Note, however, that this is a norm-dependent statement.
For example, in the commonly used~${\cal L}_2$~norm the error at shock
transitions should behave as~$O((\Delta x)^{1/2})$.)  When computations are
sufficiently well-resolved so that solutions converge as expected---that is,
\mbox{$\epsilon[\psi] \propto (\Delta x)^\alpha$,}
where $\alpha$~has its expected
asymptotic value---then they can be regarded as quite close to the $\Delta x
\!\rightarrow\! 0$~result~(\cite{finn}).  Note, however, that the Cauchy error 
exploits 
{\em no\/} knowledge of the exact continuum solution.  In practice one often
assumes that convergence at the expected rate implies convergence to the
correct solution.

Figure \newfigure\ shows that the Cauchy error measures for all three
densities~($\rho$, $m$, and~$\einternal$) decrease linearly with decreasing
gridsize~$\Delta x$.
This shows that convergence of the Cauchy error alone does {\em not\/}
imply correct behavior of the numerical scheme.
(Since the Cauchy errors for all three quantities behave
similarly, in the rest of this paper we illustrate Cauchy errors for the
momentum density only.)

Aside from the total-energy error, we have no indications that the
internal-energy scheme behaves incorrectly.  No numerical instabilities appear,
and the performance at pre-shock times is exactly as anticipated.  The error
itself is relatively small---at~$t = 40$ the error in total energy approaches
approximately~0.4\% of the initial energy in the wave---and has likely gone
unnoticed by many users of internal-energy schemes.
Since the scheme appears stable and convergent, this finite error clearly shows
that non-conservative schemes need not converge to solutions of the
conservation laws when discontinuities are present.

Motivated by this hypothesis, we make the following modification: Instead of
scaling the viscous lengths~$L_1$ and~$L_2$ with the gridsize~$\Delta x$, so
that $L_1/\Delta x$, \(L_2/\Delta x = \mbox{constants} \), we treat the viscous
lengths themselves as constants.  This approach changes the differential
equations being solved.  Viscous effects remain finite even as the
gridsize~$\Delta x$ vanishes, and so shocks are replaced by smooth transitions
with finite width.  With this change, the internal-energy scheme behaves
properly at all times.  We expect errors proportional to~$(\Delta x)^2$, since
the scheme's first-order error is entirely temporal (i.e.,~$O(\Delta t)$,
not~$O(\Delta x)$), and, by the diffusion-limiting
condition~(\ref{diff-limit}), for sufficiently small~$\Delta x$ the
timestep~$\Delta t$ is proportional to $(\Delta x)^2$.

We repeat the steepening-wave problem, now choosing $L_1 = 1/4$ and
$L_2 = 1/2$.  In Figure~\newfigure\ we see that, for
sufficiently small gridsize~$\Delta x$, the total-energy error~$\Delta E_{\it
total}\/$ tends toward zero as~$O(\Delta x)^2$ at all times.
Figure~\newfigure\ shows that the Cauchy error~$\epsilon [m]$ also
decreases as~$O(\Delta x)^2$ at all times.  The magnitudes of these errors are
much smaller than their counterparts (at the same gridsize~$\Delta x$) from the
previous ($L_1/\Delta x$, $L_2/\Delta x = \mbox{constants}$) calculations.
This suggests that these finite-gridsize results are quite close to the
solutions of the (finite-viscosity) continuum system.  Equivalently, it
suggests that the Cauchy errors measured in the previous calculations are
dominated by the change in viscous lengths~$L_1$ and~$L_2$, not by effects of
the non-zero gridsize. 

Although the preceding discussion has focussed on the failure of a particular
internal-energy scheme with particular computational parameters, the phenomenon
is more general.  We have varied the size of the timestep safety factors~$C$
and~$D$ and the strengths of the viscous parameters~$L_1/\Delta x$
and~$L_2/\Delta x$ (including setting $L_2 = 0$) with no improvement in
results.  Further, we have implemented versions with Norman and Winkler's
staggered grid, as well as several other modifications to the difference
equations~(\cite{nw}, \cite{stone}), also without improvements in results.
(Appendix~B details the modifications we have tested.)

It is possible that the scheme may converge toward the inviscid solution if we
choose to vary the gridsize~$\Delta x$ and viscous lengths~$L_1$ and~$L_2$ such
that $\Delta x, L_1, L_2 \!\rightarrow\! 0$ but~$L_1/\Delta x, L_2/\Delta x
\!\rightarrow\! \infty$.  In this situation the width of the shock
representation should decrease with increasing resolution, but the number of
gridpoints in the representation should increase.  We do not investigate the
behavior of the scheme using this procedure.
The slowly decreasing values of~$L_1$ and~$L_2$ impose penalties on
computational efficiency [via the diffusion-limiting timestep
condition~(\ref{diff-limit})],
and it is unclear whether convergence toward a physically reasonable solution
can be readily observed without choosing the viscous lengths to be nearly
constants.

To summarize, when the flow contains shocks (and the limiting equations are
inviscid), the internal-energy scheme converges toward a solution that is
demonstrably wrong.  This failure arises because the solution
contains discontinuities and the difference scheme is non-conservative.  We can
recover acceptable behavior by changing the equations so that the limiting
solution does not contain discontinuities.  In the next section we study
another test problem in which the internal-energy scheme demonstrably fails,
and also show that the total-energy scheme succeeds.


\section{Shock Tube}

Because we associate the failures of the internal-energy scheme with the
presence of shocks, for the second test problem we consider a shock tube, as
introduced by Sod~(\shortcite{sod78}).  At $t < 0$ a static, impermeable
interface separates two states of a gas at rest.  At $t = 0$ the barrier
disappears, and a shock then forms.  Because the problem admits a similarity
solution that can be found exactly, it is an ideal test problem for shocked
flows.
We test the performance of both the non-conservative, internal-energy scheme
and the conservative, total-energy scheme.

The exact solution can be constructed as follows~(\cite{landau}): Since neither
the equations of motion nor the initial conditions contain a characteristic
length, all interfaces must travel at constant speeds away from the initial
discontinuity.  A~shock propagates into the low-pressure region; a rarefaction
wave, into the high-pressure region.  The gas between these two transitions
travels at a constant speed and is uniform, except at a single point at which
the density may be discontinuous, the contact discontinuity.  The
Rankine-Hugoniot jump conditions relate the upstream velocity, pressure, and
density to the uniform middle region's conditions.  The transition across the
rarefaction wave can be obtained by noting that the flow in this region is
adiabatic, and so the physical quantities can be determined with the method of
characteristics.  Only one possible shock and one possible rarefaction wave
allow the velocity and pressure to match in the middle region; we
determine these numerically, since the relations are implicit.

With the exact solution in hand, we can test both internal-energy and
total-energy schemes.  We use Sod's initial conditions.  On the left side of
the grid, the pressure and density are given the values~$1$; on the right, the
pressure is~$1/10$ and the density is~$1/8$.  (At the interface we give these
quantities their averaged values.)  In this section only the equation of state
is given by a $\gamma = 7/5$ ideal gas.

Figure~\newfigure\ dramatically illustrates the failure of the internal-energy
scheme.  (In the shock-tube calculations, we use
\( L_1/\Delta x = L_2/\Delta x = 3/2 \).)
We show the density in the region around the shock jump at~$t = 1$ and compare
the results of the two numerical schemes with the exact solution.  Although the
computed solutions appear well-behaved and free of instabilities, two problems
with the internal-energy solution are apparent: first, the shock transition is
in the wrong place (the shock velocity is 0.2\%~too low), and second, the
post-shock density is wrong (0.3\%~too high).  Further, the total-energy
scheme gets both of these right; aside from a small post-shock oscillation, it
exhibits {\em no\/} numerical difficulties.  In Figure~\newfigure\ we show the
total-energy error accrued by the internal-energy scheme at various
resolutions, and again this diagnostic verifies that the scheme does not
converge toward a correct solution.

In Figure~\newfigure\ we show the Cauchy error
measures~$\epsilon_{\Delta x}[m]$ associated with both schemes' calculations.
For each, the errors decrease with decreasing gridsize; in fact, the quantities
are almost identical.  As before, however, we already know that the
internal-energy scheme does {\em not\/} converge to the correct solution.

The {\em rate\/} of convergence appears {\em poorer\/} than~$O(\Delta x)$.  We
trace this slow convergence to the presence of contact discontinuities, which
can present special problems to numerical schemes.  We observe the correct rate
of convergence when we alter the initial conditions to remove the
discontinuity.  This change removes much of the shock-tube problem's usefulness
as a test problem, though, since the time evolution is no longer given by a
similarity solution.
(In the next section, in which we discuss rates of convergence, we use a test
problem with smooth initial conditions that does not contain contact
discontinuities.)

Unlike shocks, contact discontinuities do not develop in smooth flow.
Similarly, the smoothed regions that represent contact discontinuities do not
tend to steepen.  Thus effects of non-zero gridsize cause them to spread
without limit, rather than toward some steady state.  For a contact
discontinuity in an isobaric medium, translating at a constant speed, no length
or time scales enter except the width~$w$ of the representation and the
gridsize~$\Delta x$.  Thus a scaling holds among~$w$,~$\Delta x$, and the
time~$t$: we may write that~$w(\Delta x, 2t) = 2 w(\Delta x/2, t)$.  It is
clear that if the representation spreads without limit, then at all
times~$w(\Delta x, 2t) > w(\Delta x, t)$ and so~$w(\Delta x/2, t) > w(\Delta x,
t)/2$.  Since the Cauchy error accrued at these discontinuities is proportional
to their widths, the Cauchy error cannot scale linearly with the
gridsize~$\Delta x$.
(We might expect that this diffusion process would lead the width~$w$
to behave asymptotically [as~$t \!\rightarrow\! \infty$] as~$t^{1/2}$.
Instead, however, using the total-energy scheme
we find that a calculation involving only the advection of contact
discontinuities suggests that asymptotically $w$~varies
as~$t^{1/3}$.)

For the shock tube, as in the steepening wave, the internal-energy scheme
appears stable and convergent but fails to converge toward a correct solution.
The Lax-Wendroff theorem shows that the conservative, total-energy scheme
should converge toward a correct solution, and all indications are that it does
so.

In brief, when the artificial-viscous lengths~$L_1$ and~$L_2$ are proportional
to the gridsize~$\Delta x$, the calculations of our internal-energy scheme do
not converge to the known shock-tube solution; while those of our total-energy
scheme do approach the exact solution.  This result reinforces our finding that
the internal-energy scheme fails in the presence of shocks.  Further, the
Cauchy errors~(\ref{cauchy}) for the (correct) total-energy scheme and the
(errant) internal-energy scheme are nearly {\em indistinguishable\/}.  Use of
convergence tests alone cannot guarantee convergence to the
correct solution.



\section{Effects of Diffusion on Convergence}

Since the internal-energy scheme is unreliable when shocks are present, we
employ the total-energy scheme exclusively for our discussion of convergence
rates.  The diffusive effects of linear artificial viscosity introduces a
noticeable amount of diffusion even to linear perturbations.  We first discuss
in some detail the impact on small, linear waves.  Then we show the effect on
convergence rates for a wall shock---a test problem containing both a shock and
small-amplitude waves.

\subsection{Linear examples}

The presence of linear artificial viscosity can have a surprisingly
large effect on the observed convergence rate.  To illustrate these effects, we
consider small-amplitude waves.  First we exhibit convergence rates for
calculations of such waves, and then show consistency between rates in
this calculation and rates derived using an analytic treatment.  Finally we
comment on the viscous lengths necessary to yield convergence rates close to
the expected asymptotic result.

For the numerical calculations, we take as initial conditions the right-moving
wave~(\ref{right-moving-wave}--\ref{modified-gaussian}) discussed above in
section~3,
with amplitude~$\delta_0 = 0.05$ and width parameter~$\Delta = 0.4$.  We
calculate to time~$t = 0.6$ using the total-energy scheme with linear
artificial viscosity only---the quadratic artificial viscous length~$L_2$
is~$0$.  Figure~\newfigure\ shows the Cauchy errors~$\epsilon [m]$.
Note that instead of plotting~$\epsilon_{\Delta x}[m]$ as a function
of~$\Delta x$ (as in Figures~2,~4, and~7), here we show it as a function of the
viscous length~$L_1$.  The computations use~$L_1/\Delta x = 4$, chosen so that
the resolution is sufficiently high that purely numerical errors (as compared
with infinitely resolved computations with those values of the viscous
length~$L_1$) are negligible.  Note that as $L_1$~decreases the decrease of the
errors tends toward the expected linear rate (cf.~the line segment
corresponding to $O(\Delta x)$~dependence), but only rather slowly.  To
quantify this tendency, for triplets of calculations we define~$\beta(L_1) =
\log(\epsilon_{2L_1}[m]/\epsilon_{L_1}[m])/\log 2$, which approximates the
logarithmic derivative of~$\epsilon_{L_1}$ with respect to~$L_1$.  In the
asymptotic regime, $\beta$~approaches the order of accuracy.  We show~$\beta$
as a function of~$L_1$ in Figure~\newfigure .  Note that even at the smallest
value of~$L_1$, $\beta$~is still rather far from the expected value of~$1$ even
though almost two orders of magnitude separate~$L_1$ from the problem's natural
length scale~$\Delta$.

A simple analytic model shows such convergence rates are a consequence of the
diffusive effects of linear artificial viscosity.  Consider the equations of
motion (\ref{euler_density}--\ref{euler_total_energy}) linearized about a
constant background, which (in terms of the small quantities~$\rho_1$,~$v_1$,
and~$P_1$) read
\begin{eqnarray}
  \frac{\partial\rho_1}{\partial t} 
     & = & - v_0 \frac{\partial\rho_1}{\partial x} 
                   - \rho_0 \frac{\partial v_1}{\partial x} \\
  \frac{\partial v_1}{\partial t}
     & = & -v_0 \frac{\partial v_1}{\partial x}
                  - \frac{1}{\rho_0} \frac{\partial P_1}{\partial x}
                  + L_1 c_0 \frac{\partial^2 v_1}{\partial x^2} \\
  \frac{\partial P_1}{\partial t}
     & = & -\gamma P_0 \frac{\partial v_1}{\partial x}
               - v_0 \frac{\partial P_1}{\partial x}.
\end{eqnarray}
(The background is specified by the constants~$v_0$,~$\rho_0$,~$P_0$,
and~\(c_0 = (\gamma P_0/\rho_0)^{1/2} \).  Note that~$L_1$ is the linear
viscous length, {\em not\/} a perturbation quantity.)  We include the
artificial-viscous terms, but to leading order in the small perturbations the
quadratic artificial viscosity vanishes.  In the {\em absence\/} of linear
artificial viscosity~($L_1 = 0$) the linearized equations take the familiar
Euler form, and the three coupled equations can be rewritten as three
{\em decoupled\/} equations,
\begin{eqnarray}
  \frac{\partial\chi_1}{\partial t} 
      & = & -v_0\frac{\partial\chi_1}{\partial x} \\
  \frac{\partial\chi_{2,3}}{\partial t}
      & = & -(v_0 \pm c_0) \frac{\partial\chi_{2,3}}{\partial x},
\end{eqnarray}
where~$\chi_1 = \rho_1 - P_1/c_0^2$
and~$\chi_{2,3} = (P_1/c_0 \pm \rho_0 v_1)/(2c_0)$, for modes that
propagate along the three characteristic paths, at speeds~$v_0$
and~$v_0 \pm c_0$.  In the {\em presence\/} of linear artificial viscosity,
performing the analogous transformation leads to three still coupled
equations:
\begin{eqnarray}
   \pderiv{\chi_1}{t} & = & -v_0 \pderiv{\chi_1}{x} \\
   \pderiv{\chi_2}{t} & = &
       -(v_0 + c_0)\pderiv{\chi_2}{x}
         + \frac{L_1 c_0}{2}\frac{\partial^2}{\partial x^2}(\chi_2 - \chi_3) \\
   \pderiv{\chi_3}{t} & = &
       -(v_0 - c_0)\pderiv{\chi_3}{x}
         + \frac{L_1 c_0}{2}\frac{\partial^2}{\partial x^2}(\chi_3 - \chi_2).
\end{eqnarray}
Note that~$\chi_1$, the mode that remains still with respect to the background
velocity~$v_0$, is unaffected by artificial viscosity.  We next {\em assume\/}
that (as in the wave example above) one of the remaining modes is sufficiently
larger than the other two so that we can neglect their influence on its
propagation.  Calling it simply~$\chi$, we thus obtain the equation of motion
\begin{equation}
 \frac{\partial\chi}{\partial t}
    + (v_0 \pm c_0)\frac{\partial\chi}{\partial x} 
      = \frac{L_1 c_0}{2} \frac{\partial^2 \chi}{\partial x^2}.
\end{equation}
The left-hand side of this equation describes just the advection of the
quantity~$\chi$, so that (with the identification of the diffusion
constant~$\nu = L_1 c_0/2$) the equation is simply the heat diffusion
equation
\begin{equation}   \label{heat}
  \frac{\partial\chi}{\partial t} = \nu \frac{\partial^2\chi}{\partial y^2},
\end{equation}
for the quantity~$\chi(y,t)$ in the frame of reference that moves at the
speed~$v_0 \pm c_0$.

We study the
convergence of solutions of the heat equation~(\ref{heat}) as the diffusion
coefficient~$\nu$
tends to
zero.  We consider the time evolution of a
Gaussian pulse.  An {\em exact\/} solution for~$t > 0$ is
\begin{equation}         \label{exact-heat}
  \chi(y,t;\nu) 
     = A \left( \frac{\Delta_0^2}{4\nu t + \Delta_0^2} \right)^{1/2} \,
         e^{-y^2/(4\nu t + \Delta_0^2)}.
\end{equation}
where~$A$ is the initial~($t = 0$) amplitude and~$\Delta_0$ is the initial
width of the pulse.  With this solution it is straightforward to calculate the
Cauchy error (at a fixed time~$t$) associated with a sequence of solutions in
which the viscous parameter~$\nu$ varies.  In Figure~\newfigure\ we show the
Cauchy error~$\epsilon_\nu [\chi]$ evaluated at~$t = 0.6$ using an
amplitude~$A = 0.05$ and width~$\Delta_0 = 0.3$.  (We choose different values
for the width parameters in this treatment and in the calculations described
above because the shape of the initial conditions differs.)  Again the
convergence rate~$\beta$, shown in Figure~\newfigure , becomes close to the
expected value of~$1$ only at quite small values of the diffusion~$\nu$.  In
the context of the flat background of the previous calculations, we require
values of the viscous length~$L_1 = 2\nu/c_0$ much smaller than the natural
length scale~$\Delta_0$ to achieve values of~$\beta$ close to~$1$.

To determine how small the diffusion coefficient~$\nu$---or, equivalently, the
viscous length~$L_1$---must be in order to achieve a convergence rate~$\beta$
close to the asymptotic value of~$1$, we examine the {\em pointwise\/}
convergence of the exact solution~(\ref{exact-heat}) of the linearized problem.
That is, for any values of~$(y,t)$, consider the behavior of~$\chi(y,t; \nu)$
as~$\nu \!\rightarrow\! 0$.  We assume that, for sufficiently small values
of~$\nu$, we may approximate
\begin{equation}    \label{error-analytic}
  \chi(y,t;\nu) - \chi(y,t; 0)
     \approx \nu \pderiv{\chi}{\nu}(y,t; 0) 
             + \frac{\nu^2}{2} \frac{\partial^2\chi}{\partial \nu^2}(y,t;0).
\end{equation}
We determine the value of~$\nu = \overline\nu (y,t)$ at which the first and
second terms on the right-hand side are equal.  The asymptotic regime for
convergence lies at values of~$\nu$ much less than~$\overline\nu$.  The
calculation itself is elementary and yields
\begin{equation}
  \overline\nu(y,t) =
      \frac{\Delta_0^2}{t}
      \left| \frac{2y^2 - \Delta_0^2}
                        {4y^4 - 12y^2 \Delta_0^2 + 3\Delta_0^4}  \right|
\end{equation}
The analogous quantity for the viscous length is
\begin{equation}  \label{l1-crit}
   \overline{L_1}(y,t) =
            \Delta_0 \frac{\Delta_0}{c_0 t}
            \left| \frac{4\Delta_0^2 y^2 - 2\Delta_0^4}
                         {4y^4 - 12y^2\Delta_0^2 + 3\Delta_0^4} \right|
\end{equation}

Loosely, the convergence rate for the Cauchy error is a weighted (spatial)
average of the pointwise convergence rate.  Since the main contribution to the
error comes from a region within a few widths~$\Delta_0$ of the spatial origin,
we replace the dimension\-less ratio in equation~(\ref{l1-crit})
by~$1/2$.
Thus, the asymptotic regime for the Cauchy error should lie at values of the
viscous length~$L_1$ much less than
\begin{equation}
   \overline{L_1} \approx \frac{\Delta_0}{2} \frac{\Delta_0}{c_0 t}.
\end{equation}
Note that {\em smaller\/} values of~$L_1$ are required as the time~$t$
increases to achieve the same convergence rate.

In practice, one may wish to increase the resolution of a computation until the
convergence rate~$\beta$ is within some tolerance~$\varepsilon$ of the
asymptotic value (here,~$1$).  Assume [as above, in
equation~(\ref{error-analytic})] that for small~$L$ the error behaves as
\begin{equation}
  \epsilon_{L} \propto \Lbar L \pm L^2,
\end{equation}
where~$\Lbar$ specifies the value of~$L$ at which the two terms on the
right-hand side contribute equally.  The logarithmic derivative of the
error~$\epsilon_L$ with respect to~$L$ is then simply
\begin{equation}
  \frac{L}{\epsilon_L} \frac{d\epsilon_L}{dL} 
     = \frac{1 \pm 2L/\Lbar}{1 \pm L/\Lbar}
     \approx 1 \pm L/\Lbar.
\end{equation}
To attain a convergence rate within~$\varepsilon$ of~$1$, we must use values
of~$L$ less than approximately~$\varepsilon\Lbar$.  (Comparison with the
results shown in Figures~\twobackfigure\ and~\samefigure\ suggest that this
estimate is about right for the analytic model and too high for the numerical
calculations by a factor of a few.)

\subsection{Wall Shock}

As noted above, even for the reliable total-energy scheme our results for the
shock-tube problem do not yield the expected $O(\Delta x)$~rate of convergence.
This is a consequence of the shock tube's discontinuous initial conditions.
For our final test problem, we choose one that contains a shock, smooth initial
conditions, and small-amplitude waves: a wall shock.

Gas impinges on a wall at~$x = 0$.  For uniform, supersonic flow from the right
($x > 0$), a shock will form and move outward.  For initial conditions we place
a transition between density, momentum, and energy values appropriate to a
Mach~2 steady-state shock.  (In the equation of state we again
use~$\gamma = 5/3$, as in the study of the nonlinear sound wave.)  We mediate
between upstream and downstream states with a smooth transition function:
\begin{eqnarray}
  v(x) & = & f(0, -9/8; x) \\
  P(x) & = & f(57/20, 3/5; x) \\
  \rho(x) & = & f(16/7, 1; x),
\end{eqnarray}
where the transition function is
\begin{equation}
  f(a, b; x) = 
   \left\{
     \begin{array}{ll}
       a, & \mbox{if $x < 3$} \\
       (a+b)/2 + [(b - a)/2]\tanh{\frac{35}{12}(x-6)}, &
                         \mbox{if $3 < x < 9$}            \\
       b, & \mbox{if $x > 9$.}
     \end{array}
   \right.
\end{equation}
At the left edge (at $x = 0$) we use reflecting
boundary conditions, but for the tests reported here we stop the calculations
at~$T = 2$ before any physical perturbations reach either edge of the
computational grid.

Although the initial conditions do not represent the exact, steady-state
profile for a shock transition, one quickly forms and translates away from the
wall.  Waves form and propagate away from the shock toward the wall.
Figure~\newfigure\ shows these structures in the density.

We have performed many calculations parameterized by the viscous length~$L_1 =
L_2 = L$ and the gridsize~$\Delta x$.  We use these to form two types of
sequences: first, with the viscous parameter~$L/\Delta x$ held constant, and
second, with the viscous length~$L$ held constant as~$\Delta x$ varies.

Figure~\newfigure\ shows the Cauchy error~$\epsilon_{\Delta x}[m]$, again as a
function of the viscous length~$L$.
Each of the five solid lines represents a sequence with
constant~$L/\Delta x$~($= 1, 2, 4, 8, 16$).  First, note that over the range of
gridsizes~$\Delta x$ in which we can reasonably calculate, the convergence rate
tends toward~$O(\Delta x)$ but does not approach it unambiguously.
Note also that the magnitude of the errors is almost entirely a function of the
viscous {\em length\/}~$L$---the solid lines overlap.  That is, the particular
choice of gridsize has only a small effect on the Cauchy error.  The difficulty
of achieving the expected convergence rate in these sequences must be an effect
of the changing values of the viscous lengths, not of the finite resolution.

We attempt to isolate the effect of the shock transition on the Cauchy error by
changing the limits of integration in its definition~(\ref{cauchy}) to include
only the transition region or to exclude it.  Figure~\samefigure\ also shows in
dotted lines the error associated with the transition region~($x > 7$), and in
dashed lines with the transients in the rest of the flow~($x < 7$).  Note that
the transition region yields the expected $O(\Delta x)$~convergence rate at
relatively large values of the gridsize~$\Delta x$.  The more slowly decreasing
contribution to the Cauchy error appears in the region of the transients.  The
Cauchy error for the shock transition should asymptotically be proportional to
the viscous length~$L$, as discussed above, and so its convergence rate implies
that transition widths are indeed scaling with~$L$.  (Appendix~C discusses the
small deviations from exact scaling in shock transition regions.)

Figure~\newfigure\ shows again Cauchy errors as a function of
gridsize~$\Delta x$.  Here, however, each solid line represents a set of
calculations with a {\em fixed\/} value of the viscous length~$L$.  We show
only calculations with the gridsize~$\Delta x$ sufficiently small that the
timestep~$\Delta t$ is always determined by the diffusion
condition~(\ref{diff-limit}); then the errors should (asymptotically) be~$
O((\Delta x)^2) $.  It is clear that these calculations lie in the asymptotic
regime: the errors are nearly proportional to~$(\Delta x)^2$.

We use the {\em same\/} calculations to construct Figures~\prevfigure\
and~\samefigure .  When considered as part of sequences with constant~$L$, they
unambiguously lie in the asymptotic regime; whereas when considered as part of
sequences with constant~$L/\Delta x$ they do {\em not.}  We illustrate this
dichotomy in Figure~\newfigure : each circle represents a calculation, and
lines represent the sequences to which each belongs.  For triplets of
calculations in sequences, we calculate the numerical convergence
rate~$\beta(\Delta x) = 
\log(\epsilon_{2\Delta x}[m]/\epsilon_{\Delta x}[m])/\log 2$, which
approaches the order of accuracy in the asymptotic regime.  Values
of~$\beta(\Delta x)$ are shown on the lines that define the appropriate
sequences of calculations.  For each \mbox{constant-$L$} sequence, where we
expect the errors to behave asymptotically as~$O((\Delta x)^2)$, we find
that~$\beta = 2$ to several decimal places.  For sequences of~$L/\Delta x =
\constant$ calculations the closest we can achieve to the expected
value~($\beta = 1$) is~$\beta = 0.93$.

The Cauchy errors in the~$L/\Delta x = \constant$ sequences are larger in
magnitude than those in the \mbox{constant-$L$} sequences (compare
Figures~\twobackfigure\ and \prevfigure ).  The smallness of the latter errors
and the attainment of the expected convergence rate shows that errors in the
former sequences are dominated by the effects of changing viscous lengths,
{\em not\/} of insufficient resolution.

The apparently poor convergence rate in the post-shock region can be related
(at least in part) to the discussion of small-amplitude waves above in
section~5.1.  The features in the post-shock region~(cf.~Figure~\backfigure{3})
consist of relatively small-amplitude waves.  A considerable contribution to
the integrand defining the Cauchy error~(\ref{cauchy}) comes from the region
near~$x = 3$, where a left-moving wave lies.  From the previous discussion, for
this wave alone we require much smaller viscous lengths~$L$ to attain a
numerical convergence rate~$\beta$ within a few percent of~$1$.  (We have
deemed this to be computationally infeasible.)

This explanation implicitly assumes that the waves generated as the shock forms
do not vary with the viscous length~$L$.  The width of the shock transition,
varies as the viscous length, however, and so some post-shock waves (also
varying as~$O(L)$) must arise from the different transition widths.  Along a
sequence in which $L/\Delta x = \constant $, these waves with short
length scale should make a contribution of order $O(\Delta x)$ to the Cauchy
error.  The effect of the other (long length scale) waves on the numerical
convergence rate continues to increase with time (i.e., $\beta$~is further
from~$1$), and so ultimately the effect of these should be greater than that of
the short-wavelength waves.

By studying this relatively simple test problem, we find that we can readily
attain the expected~$O((\Delta x)^2)$ convergence rate in sequences in which
the viscous lengths~$L_1$ and~$L_2$ are constants.  We emphasize that in such
sequences the~$\Delta x \!\rightarrow\! 0$ limit does {\em not\/} correspond to
the Euler equations~(\ref{euler_density}--\ref{euler_total_energy}).  In
sequences in which the ratios~$L_1/\Delta x$ and~$L_2/\Delta x$ are constants,
attaining the expected~$O(\Delta x)$ convergence rate appears more difficult.
The difficulty is tied to the changes in viscous lengths, not to numerical
errors introduced by non-zero gridsizes.  We attribute it to the diffusive
effects of linear artificial viscosity on waves that are generated as the shock
forms.


\section{Conclusions}

We have investigated the use of convergence tests to assess the accuracy of
finite-difference solutions to the Euler equations in the presence of shocks.
We find that a non-conservative internal-energy scheme does not converge to
solutions of the continuum equations.  Use of the Cauchy error does {\em not\/}
detect this significant failure.  The Lax-Wendroff theorem applies to our
total-energy scheme (as shown in Appendix~A), however, and so it provides
reliable results.

Our identification of this failure with the presence of shocks leads
us to introduce a modification of the equations of motion, in which the
artificial-viscous terms are held constant as the gridsize varies, rather than
scaled with the gridsize.  In such modified systems, smooth transitions
replace shock discontinuities.  With the discontinuities removed, the
internal-energy scheme converges as expected, without the indications of the
problems that signal the failure of the unmodified system.

For well-behaved schemes, convergence at the expected rate can be used as a
clear indicator of accuracy, a particularly unambiguous criterion in automated
contexts.  Diffusion due to linear artificial viscosity can lead to convergence
difficulties.  By investigating a wall shock with smooth initial conditions,
however, using our definition of Cauchy error to provide the criterion for
convergence, we readily achieve the expected convergence rate when the
artificial-viscous terms are held constant, and approach it when the
artificial-viscous terms vanish with the gridsize.

Finally, we have investigated the possibility of using the calculations that
establish our numerical convergence rate in extrapolations.  In
sequences in which the artificial viscosity scales with the gridsize, we find
that transitions that represent shocks deviate in several ways from exact
scaling.  Our findings on extrapolations and scaling can be found in
Appendix~C.


\acknowledgements

We wish to thank Sam Finn, Matt Choptuik, John Hawley, Mordecai Mac~Low, Jim
Stone, and Paul Woodward for helpful discussions.  This research has been
carried out at Cornell University with the generous support of the
NSF~(AST-8657467) under the PYI~program and NASA~(NAGW-2224) under the
LTSA~program.  PAK~wishes further to acknowledge the support of an NSF~Graduate
Fellowship.  Some computations reported herein were carried out using the
resources of the Cornell Theory Center, which receives major funding from
the~NSF and IBM~Corporation, with additional support from New York State
Science and Technology Foundation and members of its Corporate Research
Institute.

\clearpage


\appendix

\section{Lax-Wendroff Theorem, Extended}

We discuss in this appendix the extension of the Lax-Wendroff theorem to
multi-step, operator-split schemes.  This theorem was originally presented
by
Lax and Wendroff~(\shortcite{lw}); our discussion follows their treatment.

Write the system of conservation laws as
\begin{equation}       \label{pde}
  \frac{\partial u}{\partial t} = \frac{\partial f}{\partial x},
\end{equation}
where~$u$ is an unknown vector function of~$x$ and~$t$ and~$f$ is a given
vector function of~$u$.  The theorem requires that the difference equations be
written in the form
\begin{equation}            \label{update}
  v(x, t+\Delta t) = v(x,t) + \frac{\Delta t}{\Delta x} \Delta g,
\end{equation}
where~$\Delta g = g(x+\Delta x/2) - g(x-\Delta x/2)$.  The numerical flux~$g$
is determined in the following fashion: $g(x+\Delta x/2) = G(v_{-l+1},
v_{-l+2}, \ldots , v_l)$, where~$G$ is a vector-valued function of~$2l$ vector
arguments, which is related to~$f$ by the {\em single\/} requirement that
$G$~approach~$f$, in the sense that
\begin{equation}   \label{consistency}
  \lim_{\mbox{\scriptsize all } v_j \rightarrow v}
      G(v_1, v_2, \ldots , v_{2l}) = f(v).
\end{equation}
If, in the limit~$\Delta x, \Delta t \!\rightarrow\! 0$, $v(x,t)$~converges
(boundedly, almost everywhere) to some function~$u(x,t)$, then~$u(x,t)$ is a
weak solution of~(\ref{pde}).

We claim that the proof also holds if the numerical flux~$g$ is given by a
vector-valued function~$G$ of vector arguments and a (scalar)
parameter,~$\lambda = \Delta t/\Delta x$, such that
\begin{equation}         \label{consistency-extension}
  \lim_{\mbox{\scriptsize all }v_j \rightarrow v,
        \lambda \rightarrow \mbox{{\scriptsize const.}}}
       G(v_1, v_2, \ldots , v_{2l}; \lambda)           = f(v).
\end{equation}

Operator-split schemes satisfy this condition, as we now explain.  Write the
continuum equations~(\ref{pde}) as
\begin{equation}  \label{split}
  \frac{\partial u}{\partial t} = \frac{\partial \fone}{\partial x}
                                   + \frac{\partial \ftwo}{\partial x},
\end{equation}
where we take the number of substeps to be 2 (for notational simplicity).  The
difference equations will consist of two substeps to represent the equations
\begin{eqnarray}
  \frac{\partial u}{\partial t} = \frac{\partial \fone}{\partial x} \\
  \frac{\partial u}{\partial t} = \frac{\partial \ftwo}{\partial x}
\end{eqnarray}
sequentially.  That is, the first substep takes the form
\begin{equation}
  \tilde{v_j} = 
         v_j + \frac{\Delta t}{\Delta x} \Delta \gone \left( \setofv \right),
\end{equation}
where~$\setofv$ represents the set of all relevant~$v_j$.  The resulting
set~$\setofvt$ enters the second substep
\begin{equation}
  v_j^1 = \tilde{v_j} + \frac{\Delta t}{\Delta x}
                         \Delta \gtwo \left( \setofvt \right) .
\end{equation}
We assume that the substeps' numerical fluxes~$\gone$ and~$\gtwo$ are
consistent with the substep fluxes~$\fone$ and~$\ftwo$ in the sense
of~(\ref{consistency}).

The split scheme can be rewritten in the awkward one-step form
\begin{eqnarray}
  v_j^1 & = & v_j
              + \frac{\Delta t}{\Delta x} \Delta \gone \left( \setofv \right)
              + \frac{\Delta t}{\Delta x}
                \Delta \gtwo \left( \left\{ v_j 
                                   + \frac{\Delta t}{\Delta x}
                                    \Delta \gone (\setofv ) \right\} \right) \\
        & = & v_j + \frac{\Delta t}{\Delta x} \Delta 
                     \left[ \gone \left( \setofv \right)
                          + \gtwo \left( \left\{ v_j 
                                                 + \frac{\Delta t}{\Delta x}
                                                    \Delta
                                                    \gone \left( \setofv
                                                          \right)
                                         \right\}
                                   \right)
                      \right]                        \label{defines-G} \\
        & = & v_j + \frac{\Delta t}{\Delta x} \Delta {\cal G},
\end{eqnarray}
where~${\cal G}(\setofv , \Delta t/\Delta x)$ is defined by the
quantity in the square brackets.

Since the scheme can be written in the conservative form~(\ref{update}), we now
establish that the numerical flux~$\cal G$ satisfies the extended consistency
requirement~(\ref{consistency-extension}).  Let all of its vector arguments
approach the value~$v$, while~$\Delta t/\Delta x$ approaches some constant,
finite value.  Then~$\Delta \gone$ must vanish (since~$\gone$ is a continuous
function of its vector arguments where all of them are equal), and so
essentially~$\cal G$
becomes~$\gone \left( \setofv \right) + \gtwo \left( \setofv \right)$.  Since
each of~$\gone$ and~$\gtwo$ are consistent with~$\fone$ and~$\ftwo$,
then~$\cal G$ must be consistent with~$f$.



\section{Internal-Energy Scheme, Variations}

In this appendix we describe changes made to the internal-energy scheme, whose
basic form is given in section~2.  We emphasize that {\em none\/} of these
changes affects the conclusions of sections~3 and~4: the scheme still converges
to solutions that are demonstrably incorrect.

\bigskip

\noindent 1. The viscous lengths~$L_1$ and~$L_2$ and the timestep safety
factors~$C$ and~$D$ were varied in ranges given.

\begin{tabular}{ll}

parameter     &  range       \\ \hline
$L_1$         &  0.1--1      \\
$L_2$         &  1--32       \\
$C$           &  0.3--0.9    \\
$D$           &  0.008--0.9  \\

\end{tabular}

\medskip

\noindent 2. A staggered grid allows cleaner implementation of the van Leer
advection scheme.  The vector quantities---velocity and momentum---are stored
at gridpoints midway between the gridpoints for the remaining (scalar)
quantities.  This requires that every difference equation be modified: Pressure
acceleration~(\ref{pressure_acceleration}) becomes
\begin{equation} \label{pressure_acceleration_staggered}
  m_{j+1/2} \mapsto m_{j+1/2} - \frac{\Delta t}{\Delta x} (P_{j+1} - P_j).
\end{equation}
The artificial-viscosity difference equations~(\ref{art-visc-momentum},
\ref{art-visc-internal}) become
\begin{eqnarray} \label{momentum-art-visc-staggered}
  m_{j+1/2} & \mapsto &
               m_{j+1/2} - \frac{\Delta t}{\Delta x} (Q_{j+1} - Q_j) \\
  (\einternal )_j & \mapsto & (\einternal )_j - \frac{\Delta t}{\Delta x}
                                                 Q_j (v_{j+1/2} - v_{j-1/2}),
\end{eqnarray}
where now \( (\partial v/\partial x)_j = (v_{j+1/2} - v_{j-1/2})/\Delta x \).
The compressional heating equation~(\ref{comp-heat-difference}) becomes
\begin{equation}
  (\einternal)_j \mapsto (\einternal)_j - \frac{\Delta t}{\Delta x}
                                                 P_j (v_{j+1/2} - v_{j-1/2}).
\end{equation}
The van Leer advection procedure is virtually unchanged for scalar quantities.
The velocity at the staggered gridpoints is taken to
be~\( v_{j+1/2} = 2 m_{j+1/2}/(\rho_j + \rho_{j+1}) \).  For the momentum (a
vector quantity), the velocity at the other gridpoints is taken to
be~$v_j = (m_{j-1/2} + m_{j+1/2})/(2\rho_j)$.

\medskip

\noindent 3. As in Norman and Winkler~(\shortcite{nw}) and Stone and Norman~(\shortcite{stone}),
the pressure-accel\-er\-a\-tion term in the momentum
equation~(\ref{euler_momentum}) can be replaced by
\begin{equation}
  \frac{\partial v}{\partial t} = 
                                 -\frac{1}{\rho} \frac{\partial P}{\partial x}.
\end{equation}
The difference equation for the physical
pressure~(\ref{pressure_acceleration_staggered}) must be replaced by
\begin{equation}
  v_{j+1/2} \mapsto v_{j+1/2} - \frac{2\Delta t}{\Delta x}
                                     \frac{P_{j+1} - P_j}{\rho_{j+1} + \rho_j};
\end{equation}
and that for the artificial-viscosity term~(\ref{momentum-art-visc-staggered}),
by
\begin{equation}
  v_{j+1/2} \mapsto v_{j+1/2} - \frac{2\Delta t}{\Delta x}
                                     \frac{Q_{j+1} - Q_j}{\rho_{j+1} + \rho_j}.
\end{equation}

\medskip

\noindent 4. In Norman and Winkler~(\shortcite{nw}) and Stone and Norman~(\shortcite{stone}), energy
conservation is improved by using a ``time-centered pressure'' (as
in~$P^{n+1/2} = (P^n + P^{n+1})/2$) on the right-hand side of
the compressional-heating term of the internal-energy
equation~(\ref{euler_internal_energy}), 
then applying the
equation of state to replace the pressure with the internal-energy density.
The substep then can be expressed as
\begin{equation}
  (\einternal)_j \mapsto (\einternal)_j
           \frac{1 - \Delta t (\gamma - 1) (v_{j+1/2} - v_{j-1/2})/2\Delta x}
                {1 + \Delta t (\gamma - 1) (v_{j+1/2} - v_{j-1/2})/2\Delta x},
\end{equation}
which replaces~(\ref{comp-heat-difference}).

\medskip

\noindent 5. Consistent-transport advection~(\cite{nwb}) is said
to improve local conservation of certain quantities.  The advection of mass
density remains the same.  The remaining quantities are treated
``consistently'' with the mass density.  In the van Leer advection substep, the
momentum, rather than the velocity, moves the quantities over the grid.  In the
procedure for calculating the fluxes~${\cal F}_{j+1/2}$, the momentum density
field replaces the velocity field, and (to compensate for the extra factor of
mass density) the specific densities~$\psi/\rho$ replace the densities~$\psi$
themselves.



\section{Extrapolation and Transition-Region Scaling}

In the main body of this paper, we adopt the attitude that we
may regard a calculation as reliable if we can show that it lies within the
asymptotic regime (i.e., if its errors are converging as expected).  Since
such a determination uses several calculations at different gridsizes, it is
then straightforward to attempt to calculate an even more accurate solution by
extrapolating the set of calculations.

We first consider the wall-shock problem of section 5.2.  Because its physical
structure is relatively simple---one shock wave and detached waves---we
investigate the possibility of using the many calculations to perform
extrapolations to the \mbox{$\Delta x \!\rightarrow\! 0$} limit.  Where the
flow is smooth, an extrapolation in powers of the gridsize~$\Delta x$ is valid.
If the calculations extend into the asymptotic regime, then the extrapolations
should be quite accurate.

\subsection{Extrapolation at constant $L$}

First, consider extrapolations of sequences in which the viscous lengths~$L_1$
and~$L_2$ are held constant.  We use calculations for which
timesteps~$\Delta t$ are proportional to~$(\Delta x)^2$; then the truncation
error associated with each is of order~$O((\Delta x)^2)$.  We assume that, at
each gridpoint in the best-resolved calculation, the results can be expressed
as
\begin{equation}   \label{richardson}
  \psi_{\Delta x}(x, T) = \psi(x, T) + a_2(\Delta x)^2 + a_3(\Delta x)^3 +
                            \cdots,
\end{equation}
with some coefficients~$a_n = a_n (x, T)$~(\cite{richardson}).  (The linear
term is absent because the scheme is second-order.)  Extrapolation of~$N$
calculations finds the value of $\psi(x, T) = \psi_{\Delta x = 0} (x, T)$ by
eliminating the first~$N-1$ error terms.  Where values are required from
calculations at points other than gridpoints, we interpolate using polynomials
to an order of accuracy at least that of the eventual extrapolated result.  The
extrapolations are well-behaved and apparently more accurate than their
component calculations.
We illustrate this accuracy in Figure~\newfigure\ by means of a Cauchy error
obtained using extrapolations.  The extrapolations, for which~$L_1 = L_2 = L =
1/8$, use calculations (with gridsizes differing by factors of~$2$)
from~$\Delta x = 1/16$ down to the value of the gridsize indicated on the
horizontal axis.  The error decreases markedly with increasing resolution
(which also corresponds to more extensive extrapolation).

\subsection{Extrapolation at constant $L/\Delta x$}

Next, consider extrapolations in which the viscous parameter~$L_1/\Delta x =
L_2/\Delta x = L/\Delta x$ is held constant.  (For these extrapolations we
assume that the errors take a form similar to that in
equation~(\ref{richardson}), but {\em include\/} the term linear in the
gridsize~$\Delta x$, because of the first-order accuracy.)  Figure~\newfigure\
shows the density as extrapolated from a sequence with~$L/\Delta x = 2$.
(This sequence includes the calculation illustrated above in Figure~12.)
Qualitatively, the results appear poor in the region of the shock, but sensible
in the remainder of the flow.  The sharp features of the 
waves are \mbox{ungarbled} by the extrapolation, even though some features
must result from the formation of the shock discontinuity.  In the shock
region, representations of shocks approach discontinuities in
the~$\Delta x \!\rightarrow\! 0$ limit.  Hence, although the Cauchy error
measure~$\epsilon_{\Delta x} [\psi]$ for the entire flow approaches the
expected~$O(\Delta x)$ behavior, near the shock the pointwise behavior of the
error {\em cannot\/} lie in the asymptotic regime.

\subsection{Transitions in shock regions}

Although pointwise extrapolation of a sequence of calculations fails near
shocks when the viscous lengths~$L_1$ and~$L_2$ change within the sequence, one
might try to devise a special procedure for extrapolation in shock regions.
Any such method must consider the expected scaling of the
transition region with the viscous lengths. 
We investigate the mechanisms that alter this scaling.  We begin with a
steady-state problem, the
shock tube, to investigate short length scales, then move to the wall-shock
problem for the influence of large-scale gradients.

Consider a {\em steady-state\/} shock with uniform upstream and downstream
conditions.  The shock travels at constant speed, and the transition width for
the shock representation must be determined by the viscous lengths~$L_1$
and~$L_2$.  For simplicity, we take~\( L_1 = L_2 = L \).  Transitions
calculated at different viscous lengths~$L$ then are related by a scaling
relation:
\begin{equation}   \label{scaling}
  \psi(x, t; L) = \tilde\psi \left( \frac{x - x_s(t)}{L} \right)
\end{equation}
in the region of the shock, located at~$x_s$, where the function~$\tilde\psi$
is independent of~$L$.

Given two calculations~$\psi^{(1)}(x,t)$ and~$\psi^{(2)}(x,t)$ made with
different gridsizes~$\Delta x$, viscous lengths~$L$, or both, we measure the
deviation from exact scaling by means of the function
\begin{equation}    \label{deviation-function}
  D^{(1,2)}[\psi](\xi, t) =
                     \psi^{(1)}_{\Delta x^{(1)}} (x_s^{(1)} + L^{(1)}\xi, t)
                      - \psi^{(2)}_{\Delta x^{(2)}} (x_s^{(2)} + L^{(2)}\xi, t)
\end{equation}
in the vicinity of~$\xi = 0$.  Calculating this deviation function requires
knowing the shock positions~$x_s^{(1)}$ and~$x_s^{(2)}$.  In practice, when
comparing computations with different viscous lengths~$L$, we assume
that~$x_s^{(1)} = x_s^{(2)}$ and approximate the shock position by locating the
intersection of the two shock representations.  Determining~$x_s$ in other ways
does not change our results significantly.  The deviation function vanishes if
the calculations obey the exact scaling relation~(\ref{scaling}), since then
\( \psi^{(1)} (x_s^{(1)} + L^{(1)}\xi, t) =
   \psi^{(2)} (x_s^{(2)} + L^{(2)}\xi, t) = \tilde\psi (\xi) \).
When it is more
convenient to work with a scalar deviation measure, we calculate a norm
of~$D[\psi](\xi)$:
\begin{equation}    \label{deviation-scalar}
  \left| \left| D^{(1,2)}[\psi] \right| \right| =
   \int_{-\eta}^{\eta} \left| D^{(1,2)}[\psi](\xi) \right| \, d\xi .
\end{equation}
To limit the investigation to the shock region, the parameter~$\eta$ should
not be chosen too large; we take~$\eta = 5$.

We begin our discussion of deviations from scaling with the shock-tube problem,
in which the shock is truly steady-state.  Our comparisons of calculations fall
into two categories: first, we compare calculations with the {\em same\/}
viscous length~$L$; and, second, we compare calculations with {\em different\/}
values of~$L$.  In the second type, comparisons between calculations with the
{\em same\/} value of the ratio~$L/\Delta x$ are a useful special case.  (For
illustrations we focus on the sequence with~$L/\Delta x = 4$.)

Before describing the causes of scaling deviations, we first discuss
some useful properties of shock-tube calculations.  In this problem the only
length scale set by initial conditions and equations of motion is the
artificial-viscous length~$L$.  Thus solutions of the continuum equations
(including the effects of artificial viscosity) obey the scaling property
\begin{equation}
  \psi (x,t; L) = \psi (\sigma x, \sigma t; \sigma L),
\end{equation}
with~$\sigma$ arbitrary, where the spatial origin is chosen at the point of the
initial discontinuity.  The corresponding finite-difference solutions have an
analogous property.  A difference solution~$\psi_j^n = \psi_{\Delta x} (j\Delta
x, t_n; L)$, which approximates the continuum solution~$\psi_L (j\Delta x,
t_n)$, may equally well be interpreted as the result of a calculation in which
the values of the viscous length~$L$, the gridsize~$\Delta x$, and the
timesteps~$\Delta t$ are all scaled by~$\sigma$: $\psi_j^n = \psi_{\sigma
\Delta x} (j\sigma\Delta x, \sigma t_n; \sigma L)$.  Hence, two calculations
with the same value of the viscous parameter~$L/\Delta x$ made to the same
time~$T$ at different gridsizes~$(\Delta x)_1$ and~$(\Delta x)_2$ are
completely equivalent to two calculations with the same~$L/\Delta x$
and~$(\Delta x)_1$, but made to the times~$T$ and~$( (\Delta x)_1 / (\Delta
x)_2 ) T$, respectively.

\subsubsection*{Truncation error (shock tube)}

We have identified several causes of deviations from exact scaling.  First, if
the two viscous lengths entering into~$D[\psi]$ are the
same, then actually it does not measure deviations from scaling.  Instead, it
measures the truncation errors proportional to~$(\Delta x)^2$ of computed
solutions (as compared to the ``exact'' solution with that same {\em finite\/}
value of~$L$).

We estimate the magnitude of these truncation errors in a calculation at a
given gridsize~$\Delta x$ by differencing it with a calculation at the smaller
gridsize~$\Delta x/2$ (with the same value of the viscous length~$L$).  In
Figure~\newfigure\ we show this estimate of truncation errors as a function of
the scaled spatial co\"ordinate~$\xi = (x - x_s)/L$ in the shock-tube problem
for calculations (to the fixed time~$T = 1$) at gridsizes~$\Delta x = 1/100,
1/200, 1/400, 1/800$, all with the viscous length~$L$ chosen so that~$L/\Delta
x = 4$.  Note that the results are all quite similar despite the different
values of~$L$ and~$\Delta x$.  Along a sequence in which~$L/\Delta x =
\constant$, the effect of truncation error is a constant in the scaled
co\"ordinate~$\xi$.

We explain this result as follows: Interpret the results as a sequence in
{\em time\/} of truncation errors at a single gridsize and viscous length.  If
this chosen gridsize is~$\Delta x = 1/800$, then Figure~\samefigure\ shows
results at times~$t = 1/8, 1/4, 1/2, \mbox{and } 1$, respectively.  It then
indicates that when the transition is near steady-state, the truncation error
remains nearly constant in time.

\subsubsection*{Time-dependent relaxation (shock tube)}

The solid line in Figure~\newfigure\ shows the deviation function~$D[\rho]$
that compares transitions calculated with viscous lengths~$L^{(1)} = 1/25$
and~$L^{(2)} = 1/50$, with the gridsize chosen sufficiently small~($\Delta x =
1/1600$) so that the $O((\Delta x)^2)$~truncation errors are small.  This
illustrates a second source of deviations from exact scaling.  We identify it
as an approach to steady state, as the discontinuous initial conditions relax
toward a steady-state computed transition.  (The shape of the functions in
Figure~\samefigure\ show that transitions are steeper than the idealized,
scaling case.)  This relaxation is a consequence of the artificial viscosity,
and so it is reasonable that its time scale must be on the order of~$L/u$
(where~$u$ is some characteristic velocity), which is the only time scale
available in the problem.  (Alternatively, we may argue that
the diffusion time scale for a length~$\ell$ is on the order of~${\ell}^2/{\cal
D}$, with diffusion coefficient \mbox{${\cal D} \propto uL$}.  Since the
computed transition width evolves from zero to a value on the order of~$L$, the
resulting time scale is on the order of~$L/u$.)  When the ratio~$L/T$ is
smaller, the transition should be closer to steady state [i.e., the
exact-scaling transition function~(\ref{scaling})].

Figure~\samefigure\ also shows (dashed line) the deviation function calculated
with the same values of the viscous length~$L$ but with larger gridsizes,
chosen with the same value of the ratio~$L/\Delta x = 4$.  With such large
gridsizes, the~$O((\Delta x)^2)$ truncation error associated with either
calculation is much larger in magnitude than the deviation function.  The
deviation function, however, is nearly identical to that in the more accurate,
small-gridsize calculations.  This shows a consequence of the similarity in
truncation errors as a function of~$\xi = (x - x_s)/L$ along sequences with
fixed~$L/\Delta x$:  Write the result of a calculation at gridsize~$\Delta x$
and viscous length~$L$ in the vicinity of the shock as
\begin{equation}       \label{truncation-error}
  \psi_{\Delta x} (x,t; L) =
             \psi (x,t; L)
             + {\cal T} \left( \frac{x-x_s}{L}, t; \frac{L}{\Delta x} \right) ,
\end{equation}
where the truncation error~$\cal T$ depends on~$\Delta x$ through the
ratio~$L/\Delta x$, as discussed above.  Then the deviation function for
calculations with the {\em same value\/} of~$L/\Delta x$ reduces to
\begin{eqnarray}
  D^{(1,2)}[\psi](\xi) & = & \psi^{(1)} (x_s + L^{(1)}\xi, t; L^{(1)})
                        + {\cal T} \left( \xi, t; \frac{L}{\Delta x} \right) 
                                                            \nonumber       \\
                       &   &
                  \mbox{  } - \psi^{(2)} (x_s + L^{(2)}\xi, t; L^{(1)})
                        - {\cal T} \left( \xi, t; \frac{L}{\Delta x} \right) \\
                       & = & \psi^{(1)} (x_s + L^{(1)}\xi, t; L^{(1)})
                         - \psi^{(2)} (x_s + L^{(2)}\xi, t; L^{(1)}), \nonumber
\end{eqnarray}
that is, the deviation function for calculations without truncation errors.

We use this result to determine~$\Delta x \!\rightarrow\! 0$ deviation
functions over a wide range of viscous lengths~$L$, including values of~$L$ for
which sufficiently accurate calculations (i.e., with small truncation errors)
are computationally infeasible.  In Figure~\newfigure\ we show the scalar
deviation measure~(\ref{deviation-scalar}) as a function of~$L$, determined
from a sequence of calculations at gridsizes~$\Delta x = 1/100, 1/200, \ldots ,
1/3200$, with~\mbox{$L/\Delta x = 4$}.  In the reinterpretation of these
calculations as a sequence in increasing {\em time\/} (rather than increasing
{\em resolution\/}), we have a sequence at times~$t = 1/32, 1/16, \ldots , 1$.
Because the deviations result from a diffusive relaxation process, we na\"ively
expect their amplitudes to decrease exponentially with increasing time (or
decreasing~$L$).  Clearly the deviations decrease, but not exponentially.  (The
behavior appears more similar to a power law, but it does not approach one
asymptotically.  In any case, we have no reason to expect such behavior.)

Morduchow and Paullay~(\shortcite{mord71}) and Morduchow and
Valentino~(\shortcite{mord79}), working with a one-dimensional ideal gas with a
Prandtl number of 3/4, performed an eigenvalue analysis to investigate shock
stability.  They found a continuum of decay rates, including values arbitrarily
close to zero.  If our system shares this property, then even asymptotically we
should not expect to observe exponential decay.  A combination of perturbation
modes with a continuum of rates could imitate the observed behavior.

\subsubsection*{Large-scale gradients (wall shock)}

Our third and last cause of deviations from exact scaling has a simple physical
origin.  Large-scale gradients in the flow have an effect on the structure of
the transition.  Since transitions of different widths sample these gradients
to different extents, observable deviations result.  To illustrate, we
consider a physical problem in which the shock itself is not in steady state,
the wall-shock problem.  Although as~$t \!\rightarrow\! \infty$ the
shock approaches steady state, at any finite time the shock generates weak
gradients in the downstream region.

Figure~\newfigure\ shows deviation functions~$D[\rho](\xi)$ for the wall-shock
problem.  The two curves represent comparisons of, first (solid line),~$L^{(1)}
= 1/128$ and~$L^{(2)} = 1/256$ (using~$L/\Delta x = 8$), and second (dashed
line),~$L^{(1)} = 1/256$ and~$L^{(2)} = 1/512$ (using~$L/\Delta x = 4$).  (The
sense of the deviation shows that the transition is steepening; this agrees
with the smoothed initial conditions, which [by design] are less steep than the
steady-state transition.)  Note the effect of the large-scale gradients
downstream from the shock---unlike in the steady-state, shock-tube cases,
$D(\xi)$ tends toward non-zero values for negative values of~$\xi = (x -
x_s)/L$.  The slope (with respect to~$\xi$) with which the curve turns away
from zero on the left side should be proportional to the scaling factor,
i.e.,~$L$, as Figure~\samefigure\ verifies.

\subsection{Extrapolation difficulties}  

Our investigations into extrapolation procedures suggest difficulties in
devising an automated extrapolation scheme that ensures a predetermined order
of accuracy.  Extrapolating sequences with the artificial-viscous length~$L$
fixed appears feasible, but the improvements are small.  Direct
pointwise extrapolation of sequences in which the viscous length~$L$ is chosen
proportional to the gridsize~$\Delta x$ yields improvements away from shocks,
but at the cost of reasonable behavior at shocks.  Finally, our studies of the
scaling behavior of shock transitions suggests that computed transitions in
general deviate from exact scaling behavior in ways not easily
controlled; exploiting the scaling behavior to extrapolate to zero-width shock
transitions may incur errors of unknown magnitude.

\clearpage



\section*{Figure captions}

\begin{captions}

\item[1.] Difference between total energy in an internal-energy computation and
exact total energy, as a function of gridsize~$\Delta x$ in the steepening-wave
problem of section~3.  Artificial viscosity vanishes with decreasing gridsize.
Each line corresponds to a different time (from bottom to top, $t = 4, 8, 12,
\ldots, 40$).  When $t > 24$, the error does not tend toward zero.

\item[2.] Cauchy errors~(\ref{cauchy}) for momentum density in the
steepening-wave problem.  The two lines correspond to $t = 20$ and $t = 40$.

\item[3.] Difference between total energy in an internal-energy computation and
exact total energy, as a function of gridsize~$\Delta x$ in the steepening-wave
problem.  The viscous lengths~$L_1$ and~$L_2$ are fixed.  Each line corresponds
to a different time ($t = 4, 8, 12, \ldots, 40$).

\item[4.] Cauchy errors for the steepening-wave problem.  The viscous
lengths~$L_1$ and~$L_2$ are fixed.  The two lines correspond to $t = 20$ and 
$t = 40$.

\item[5.] The region near the shock front in the shock-tube problem
calculated by both internal-energy (solid curve) and total-energy schemes
(dashed curve) (at~$t = 1$, using~$\Delta x = 1/6400$
and~$L_1 = L_2 = 3/12800$) as well as the exact solution (dotted curve).

\item[6.] Difference between total energy in an internal-energy computation of
the shock tube and the exact total energy, as a function of
gridsize~$\Delta x$.  Increased resolution does not yield a better result.

\item[7.] Momentum Cauchy errors for both internal-energy and total-energy
calculations of the shock tube.  Their behaviors are almost indistinguishable,
although the internal-energy calculation is wrong.

\item[8.] Momentum Cauchy errors for a small-amplitude sound wave.  The
quadratic viscous length~$L_2$ is zero.  The linear viscous length~$L_1$ is
chosen to be~$4\Delta x$.  The short line segment indicates a slope
corresponding to~$\Delta x$ dependence.

\item[9.] The numerical convergence rate~$\beta$ for a small-amplitude sound
wave.  The viscous lengths are~$L_1 = 4\Delta x$ and~$L_2 = 0$.

\item[10.] Cauchy errors for the analytic solution of the heat equation as
diffusion coefficient~$\nu \!\rightarrow\! 0$.  As discussed in the text, the
linearized equations of motion in the presence of linear artificial viscosity
may take this form.

\item[11.] The numerical convergence rate~$\beta$ associated with solutions of
the heat equation as~$\nu \!\rightarrow\! 0$.

\item[12.] Density as a function of~$x$ for the wall-shock problem of
section~5, at~$t = 2$, computed with~$L_1 = L_2 = 1/128$ at~$\Delta x = 1/256$.

\item[13.] Momentum Cauchy errors for the wall-shock problem.  Each solid
line describes a sequence of calculations in which the viscous length~$L_1 =
L_2 =L$ is a constant multiple of the gridsize; five lines represent the
values~$L/\Delta x = 1, 2, 4, 8, 16$.  The short line segment indicates a slope
corresponding to~$\Delta x$ dependence.  The dashed and dotted lines isolate
the contributions from the shock-transition region ($x < 7$) and the rest of
the flow~($x > 7$), respectively.

\item[14.]  Momentum Cauchy errors for the wall-shock problem.  Each line
describes a sequence of calculations in which the viscous lengths~$L_1 = L_2 =
L$ are held constant (from left to right, $L = 1/256, 1/128, \ldots, 1/8$).

\item[15.] Convergence of the momentum Cauchy errors for the wall-shock
problem, considered within sequences in which the viscous lengths~$L_1$
and~$L_2$ are {\em constant\/} (horizontal lines) and {\em proportional to the
gridsize~$\Delta x$\/} (diagonals).  Numerical convergence rates~$\beta$, shown
along these sequences, measure the approach to the expected convergence
rates~$O(\Delta x)$ and~$O((\Delta x)^2)$.

\item[16.] Momentum Cauchy errors for extrapolations in the wall-shock problem,
with $L = 1/8$.  The extrapolations use calculations from~$\Delta x = 1/16$
down to the value of the gridsize shown on the horizontal axis.

\item[17.] The result of a pointwise extrapolation of three calculations of the
wall-shock problem, at
gridsizes~$\Delta x = 1/64, 1/128, 1/256$, with~$L/\Delta x = 2$.

\item[18.] The~$O((\Delta x)^2)$ truncation error (estimated as discussed in
text) as a function of~$\xi = (x - x_s)/L$ in the shock-tube problem
at~$t = 1$.  The estimates are for four calculations, all
with~$L/\Delta x = 4$, with gridsizes~$\Delta x = 1/100, 1/200, 1/400, 1/800$.
As discussed in the text, these calculations may also be interpreted as a
sequence in time~($t = 1/8, 1/4, 1/2, 1$) at~$\Delta x = 1/800$.

\item[19.] Deviation from exact shock-transition scaling of
density~(\ref{deviation-function}) as a function of $\xi = (x - x_s)/L$ in the
shock-tube problem.  The solid line compares~$L = 1/25$ and $L = 1/50$
calculations; the $O((\Delta x)^2$~truncation error is negligible.  The dashed
line compares~$L = 1/25$,~$\Delta x = 1/100$ and~$L = 1/50$,~$\Delta x = 1/200$
calculations; the $O((\Delta x)^2)$ truncation error for either is
larger in magnitude than the function plotted here.

\item[20.] Deviation from exact shock-transition scaling of
density~(\ref{deviation-scalar}) as a function of viscous length~$L$.
The~$O((\Delta x)^2)$ truncation error has been removed, as described in the
text, leaving the effects of relaxation toward exact scaling.

\item[21.] Deviation from exact shock-transition scaling of
density~(\ref{deviation-function}) as a
function of \mbox{$\xi = (x - x_s)/L$} for the wall-shock
problem.  The solid curve represents a comparison
of~$L = 1/128$ with~$L = 1/256$; the dashed curve, of~$L = 1/256$ with
\mbox{$L = 1/512$}.

\end{captions}

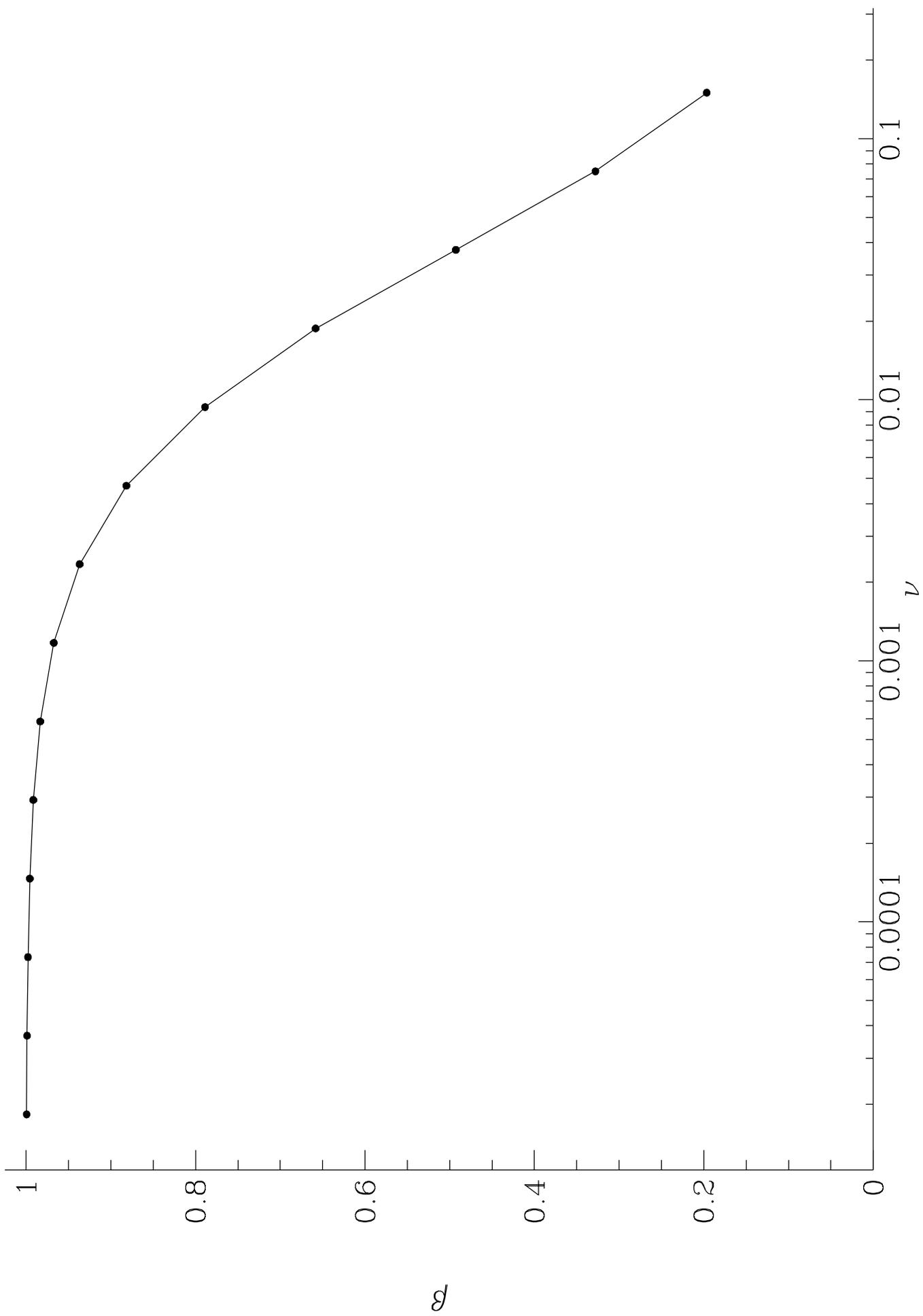
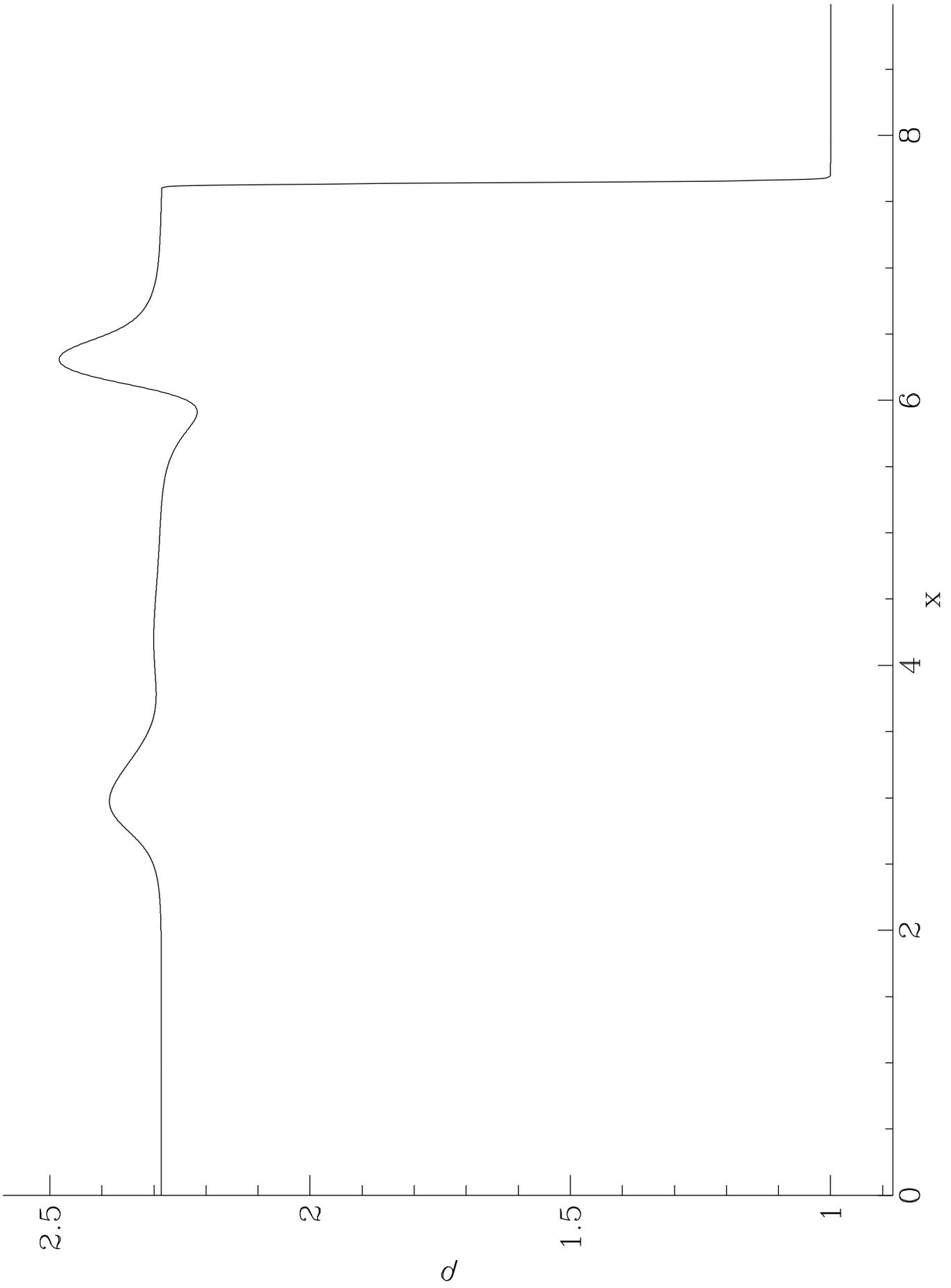
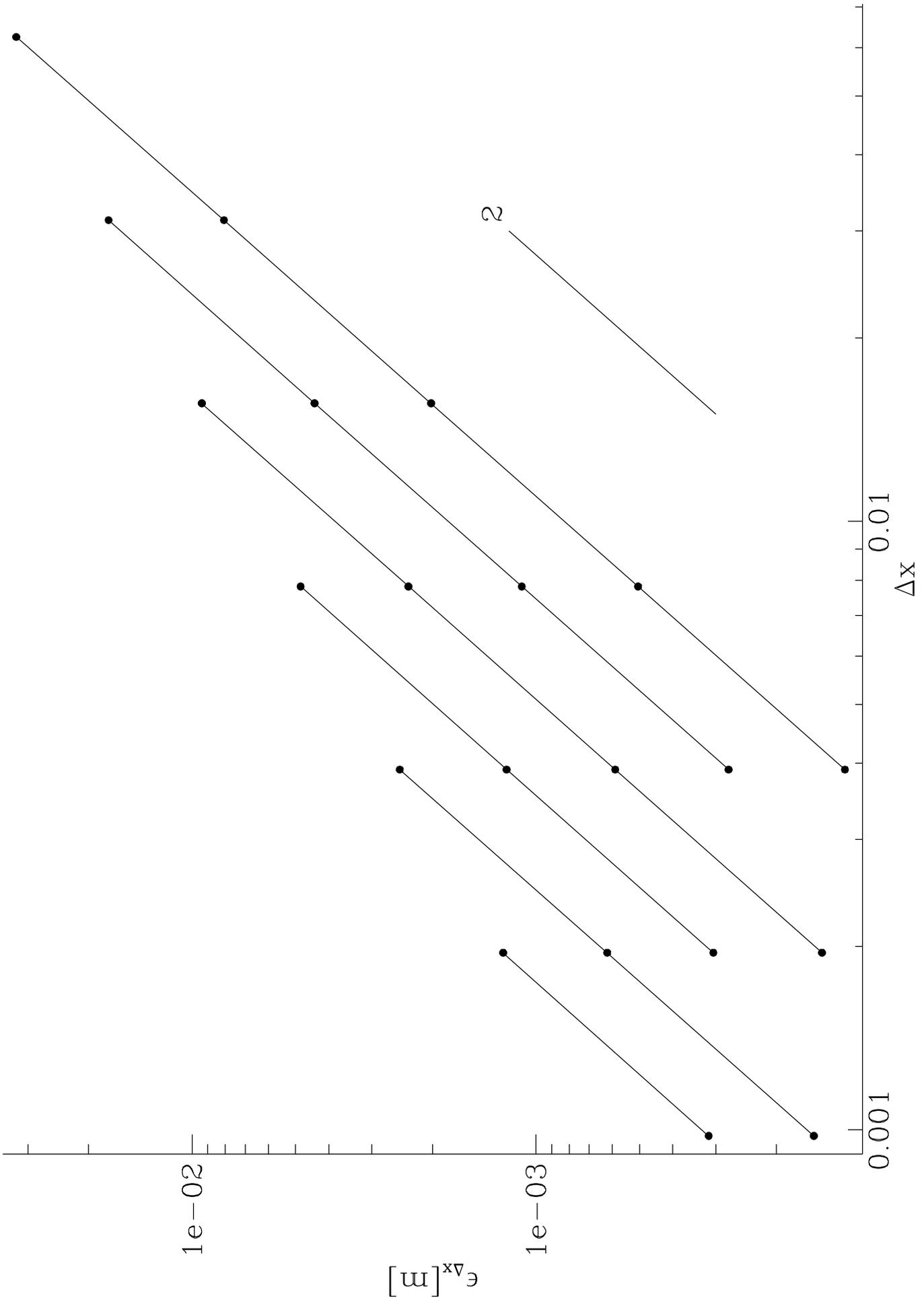
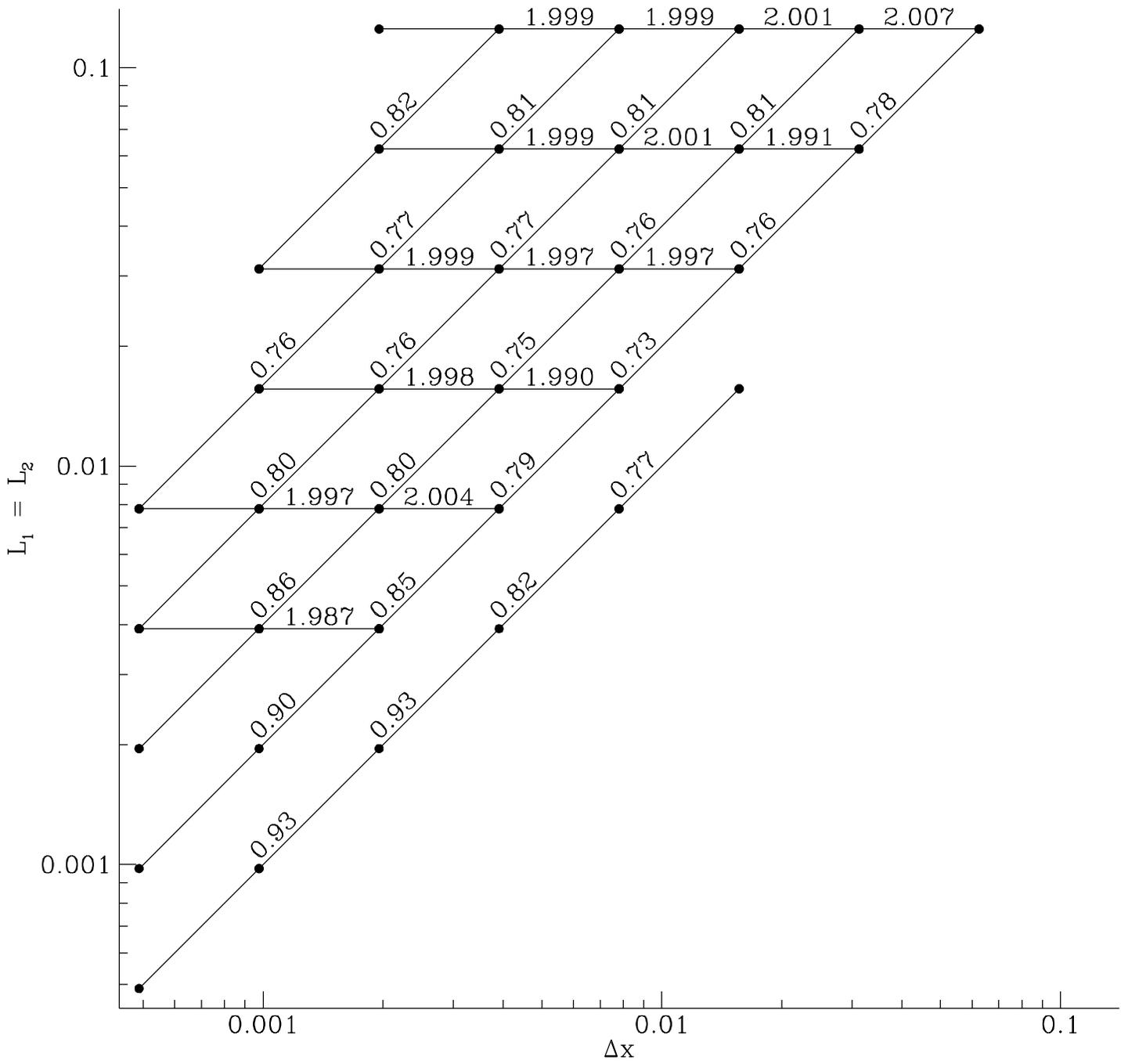
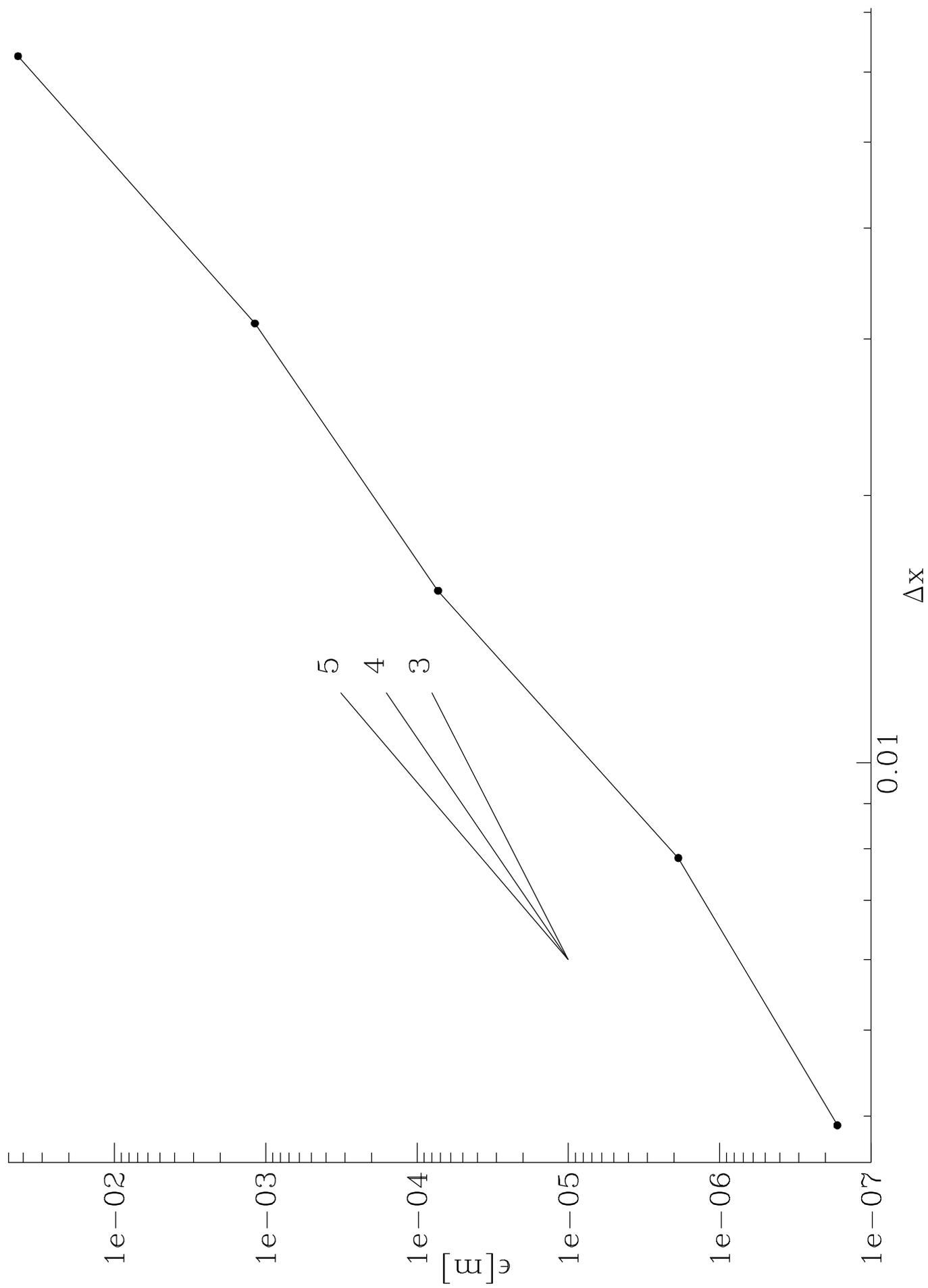
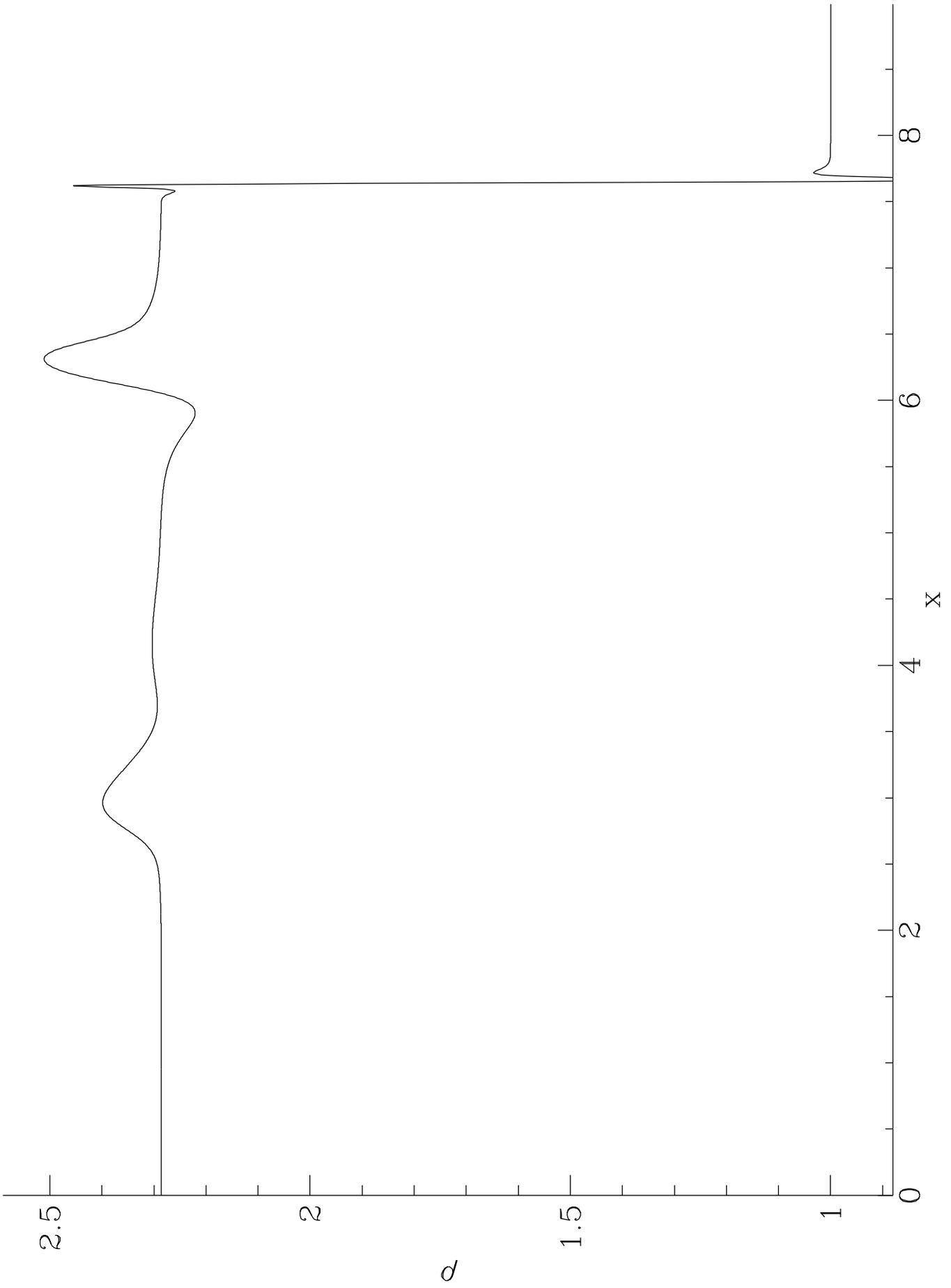
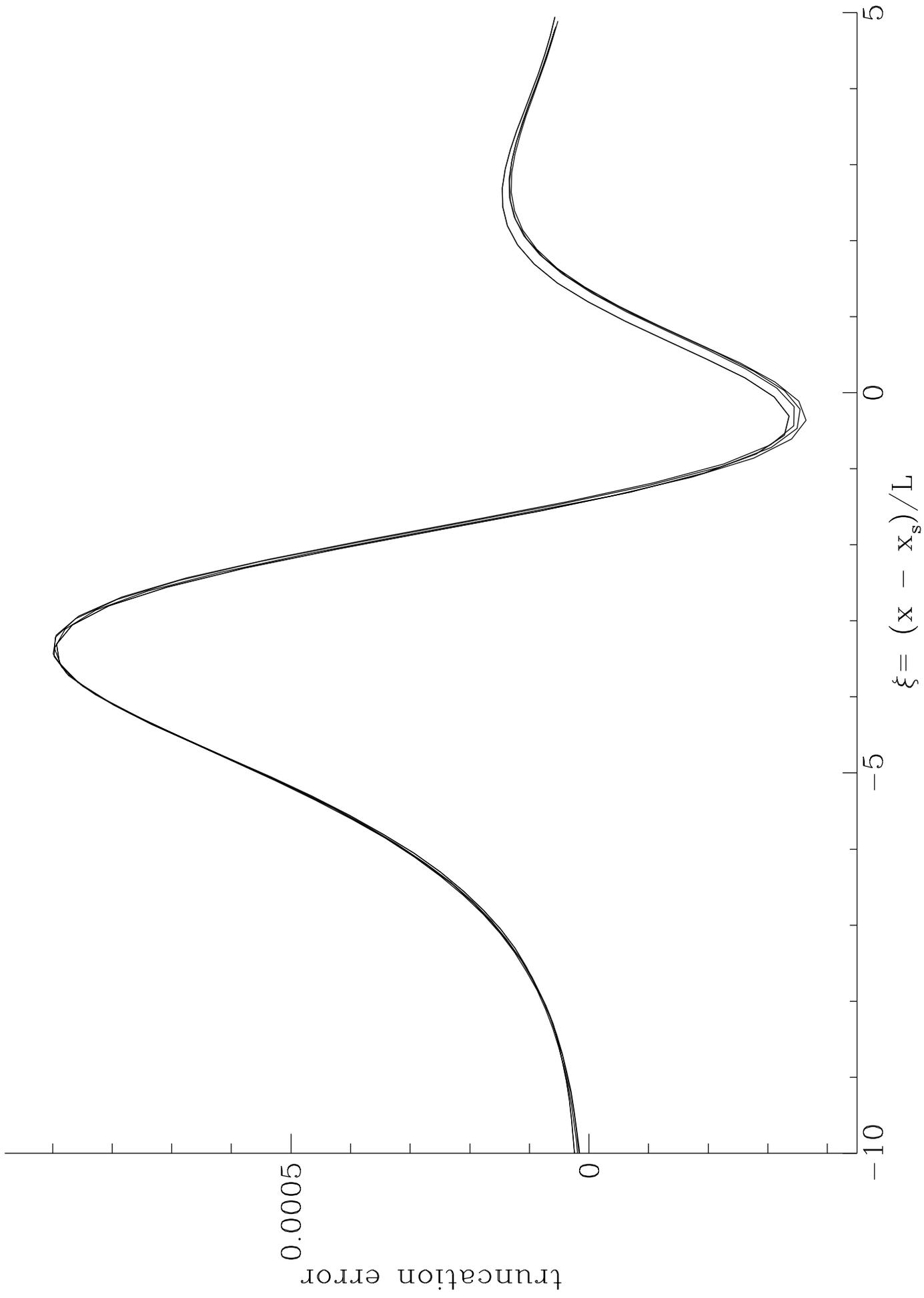
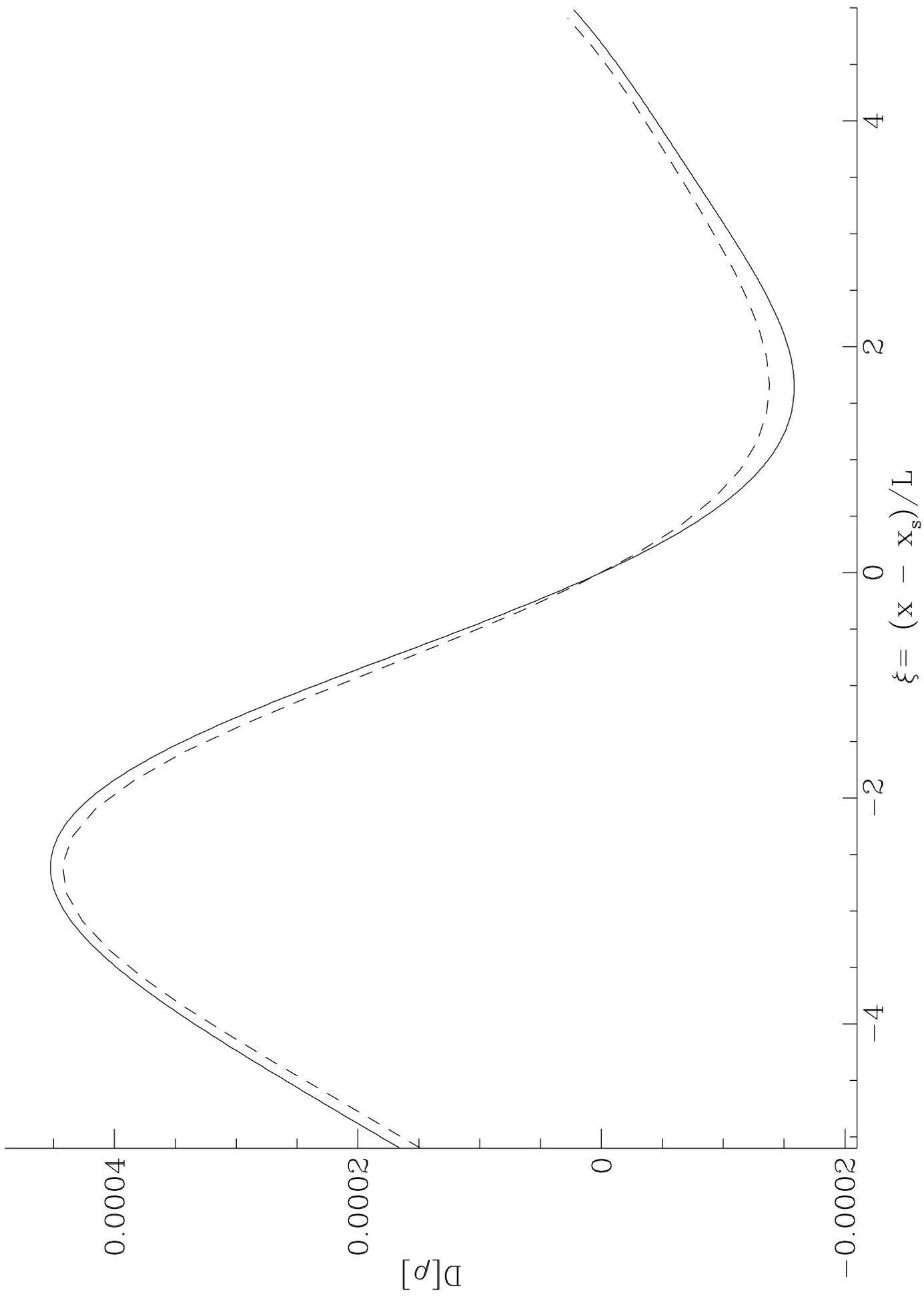
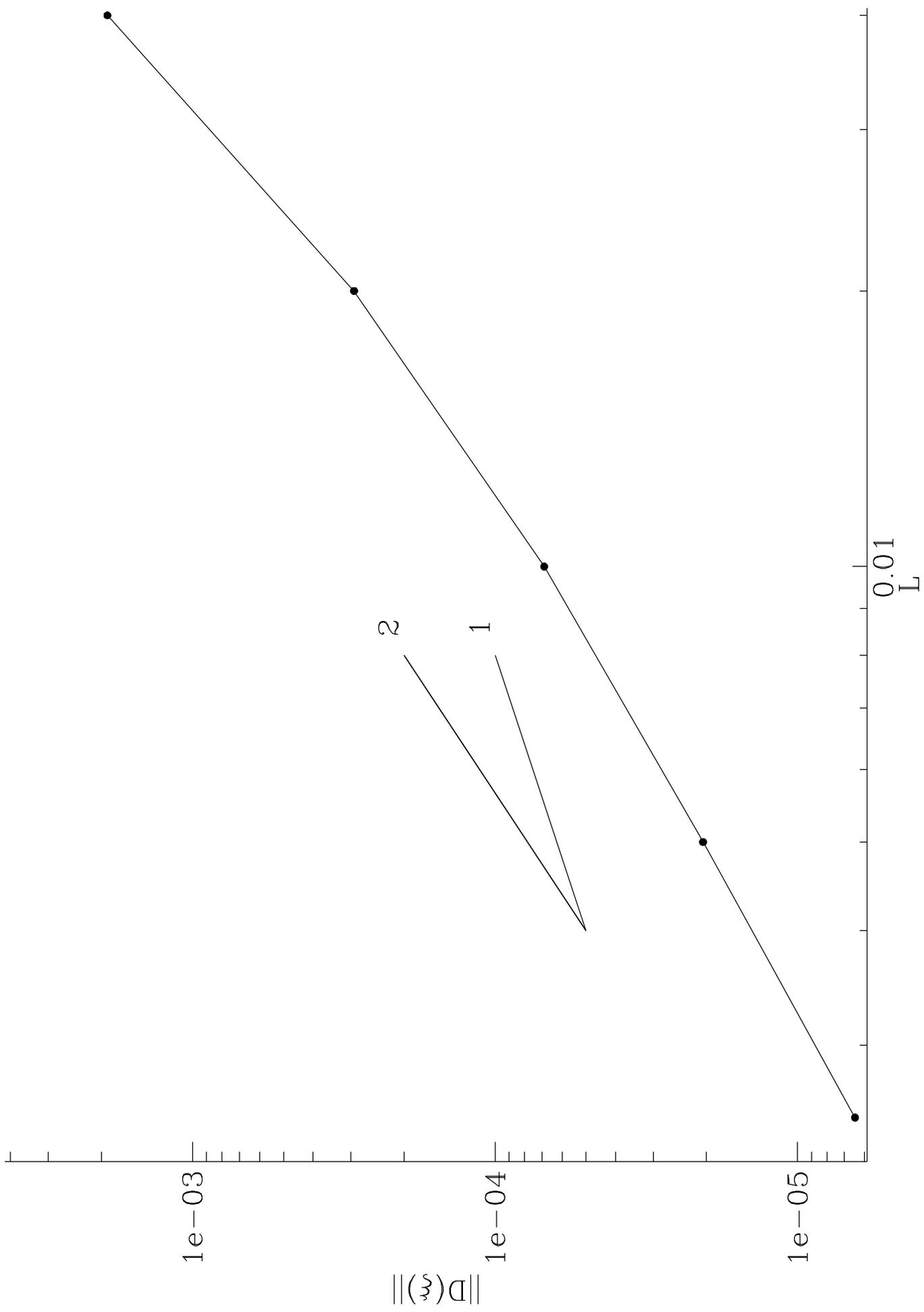
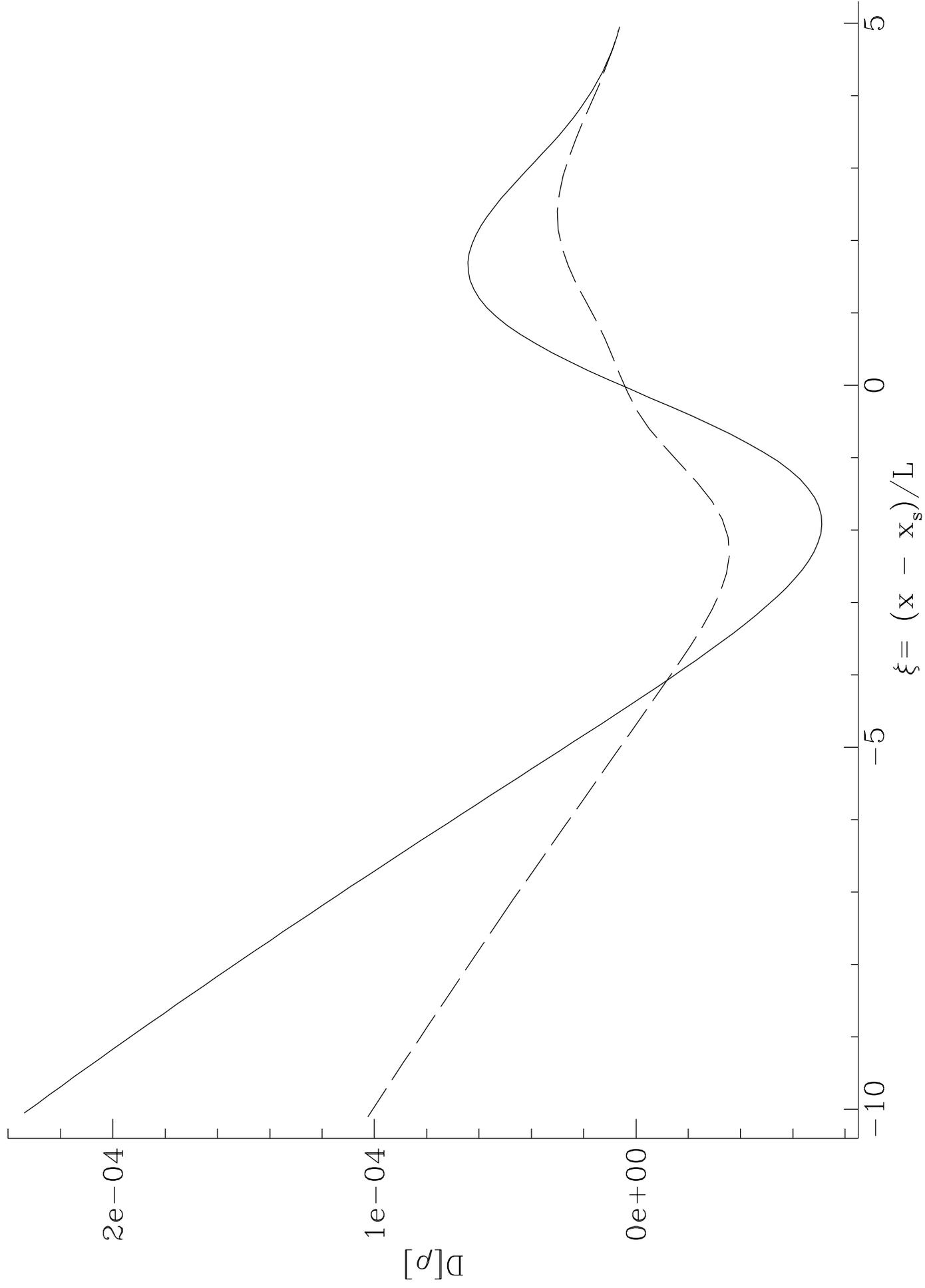
\end{document}